\documentclass[floats,floatfix,amssymb,prd,twocolumn,superscriptaddress,nofootinbib,preprintnumbers]{revtex4-1}

\usepackage{ragged2e}
\usepackage{amssymb,amsmath,verbatim,mathtools,needspace,enumitem,etoolbox,graphicx,microtype,afterpage,bm}

\usepackage{tikz}
\usetikzlibrary{shapes.misc, positioning, arrows.meta}
\usepackage{framed}

\usepackage{hyperref}
\hypersetup{colorlinks, citecolor=bluscuro, linkcolor=black, urlcolor=bluscuro}
\definecolor{rossos}{cmyk}{0,1,1,0.55}
\definecolor{bluscuro}{rgb}{0.15, 0.2, .85}
\definecolor{bluchiaro}{cmyk}{1,.3,0.,0.1}
\definecolor{ForestGreen}{rgb}{0.13, 0.55, 0.13}
\definecolor{TLGreen}{RGB}{50, 164, 49}
\definecolor{TLOrange}{RGB}{231,180,22}
\definecolor{TLRed}{RGB}{204,50,50}

\usepackage{multirow,pifont,lmodern,float,tabularx,booktabs}
\usepackage[all]{hypcap}
\usepackage[T1]{fontenc}
\usepackage[utf8]{inputenc}

\renewcommand{\arraystretch}{1.4}

\allowdisplaybreaks

\newcommand{\be}{\begin{equation}}
	\newcommand{\ee}{\end{equation}}
\renewcommand{\d}{{\rm d}}

\newcommand{\tc}{t_\textrm{c}}
\newcommand{\zetapt}{\zeta_\textrm{PT}}

\newcommand{\cern}{
	CERN, Theoretical Physics Department,
	Esplanade des Particules 1, Geneva 1211, Switzerland}

\newcommand{\IAP}{Institut d'Astrophysique de Paris, UMR 7095 du CNRS et de Sorbonne Universit\'e,\\ 98 bis bd Arago, 75014 Paris, France}

\begin{document}

	\title{Ultra-Slow-Roll Inflation on the Lattice II:\\  Nonperturbative Curvature Perturbation}
	\author{Angelo Caravano}
	\email{caravano@iap.fr}
	\affiliation{\IAP}
	
	\author{Gabriele Franciolini}
	\email{gabriele.franciolini@cern.ch}
	\affiliation{\cern} 
	
	\author{Sébastien Renaux-Petel}
	\email{renaux@iap.fr}
	\affiliation{\IAP}
	
	\date{\today}
	
	\begin{abstract}
		\noindent
		Building on the recent lattice simulations of ultra-slow-roll (USR) dynamics presented in~\cite{Caravano:2024moy}, we investigate the role of the nonlinear relation between the inflaton field configuration and the curvature perturbation $\zeta$, the key observable after inflation. Using a nonperturbative $\delta N$ approach applied to the lattice output, we generate fully nonlinear three-dimensional maps of $\zeta$. This calculation captures both the non-Gaussianity arising from the nonlinear mapping between $\phi$ and $\zeta$, and the intrinsic non-Gaussianity generated around Hubble crossing by the nonlinear field dynamics, which is neglected in stochastic approaches. We find that the nonlinear mapping has a profound impact on the statistics, significantly enhancing the positive tail of the $\zeta$ probability distribution, with important implications for observable quantities. A central part of this work is the comparison with the standard perturbative treatment based on a gauge transformation, which allows us to quantify
		when and how the perturbative picture breaks down as fluctuations grow large. Together with~\cite{Caravano:2024moy}, this work sets the basis for robust, nonperturbative predictions of primordial black hole production and scalar-induced gravitational wave emission from inflation using lattice simulations.
	\end{abstract}
	
	\preprint{CERN-TH-2025-116}
	
	\maketitle
	
	{
		\setcounter{tocdepth}{1}
		\hypersetup{linkcolor=black}
		\tableofcontents
	}
	\hypersetup{linkcolor=bluscuro}

	\section{Introduction}\label{intro}
	Slow-roll inflation is widely accepted as the simplest
	mechanism for explaining the origin of cosmic fluctuations. Its predictions are in excellent agreement with observed large-scale properties, such as those of the cosmic microwave background (CMB) \cite{Akrami:2018odb,Planck:2019kim,BICEP:2021xfz}. In the late stages of the inflationary evolution, however, when fluctuations on smaller scales cross the Hubble scale, a departure from the slow-roll regime can significantly amplify perturbations, leading to various observable signatures.
	These are, among others, the emission of scalar-induced gravitational waves (SIGWs) \cite{Tomita:1975kj, Matarrese:1993zf, Acquaviva:2002ud, Mollerach:2003nq, Ananda:2006af, Baumann:2007zm, Domenech:2021ztg} and the formation of primordial black holes (PBHs) \cite{Zeldovich:1967lct,Hawking:1971ei,Carr:1974nx,Carr:1975qj,Chapline:1975ojl}. These small-scale observables offer a complementary probe of the inflationary dynamics, potentially revealing features that are inaccessible through observations of the CMB and the large-scale structure of the universe.
	While a broad range of models can produce such effects, a large amplification of fluctuations challenges the validity of perturbation theory, making it challenging to obtain reliable predictions.
	
	This has motivated ongoing efforts in the literature to extend inflationary predictions beyond perturbation theory using simulations~\cite{Caravano:2021pgc,Caravano:2021bfn,Caravano:2022epk,Caravano:2022yyv,Figueroa:2023oxc,Caravano:2024xsb,Caravano:2024tlp,Mizuguchi:2024kbl,Launay:2024qsm,Florio:2024pgm,Animali:2025pyf,Launay:2025kef}. In particular, lattice simulations—widely used to study (p)reheating and cosmological phase transitions—are emerging as a crucial tool in this context~\cite{Caravano:2021pgc,Caravano:2021bfn,Caravano:2022epk,Caravano:2022yyv,Caravano:2024xsb,Figueroa:2023oxc,Caravano:2024tlp}. In \cite{Caravano:2024moy} (hereafter Paper I), we performed lattice simulations of inflationary models featuring an ultra-slow-roll (USR) phase, a well-known mechanism for enhancing fluctuations through a transient deceleration of the inflaton \cite{Ivanov:1994pa,Kinney:1997ne,Inoue:2001zt,Kinney:2005vj,Martin:2012pe,Motohashi:2017kbs}. Paper I focused on the dynamics of the inflaton, highlighting a backreaction effect that modifies the background evolution, as well as measuring the deviations from Gaussianity of the field perturbations induced by self-interactions. However, making quantitative predictions requires not only tracking the inflaton dynamics but also computing the resulting observables after inflation, such as the comoving curvature perturbation $\zeta$.
	
	In this work, we extend the analysis of Paper I by extracting the fully nonperturbative curvature perturbation \(\zeta\) from our lattice simulations and studying its properties, such as its power spectrum, its clustering and its probability density function, including rare large fluctuations.
	Our method naturally incorporates both the nonlinear effects generated around horizon crossing, which are not taken into account in stochastic approaches (see e.g. \cite{Ezquiaga:2019ftu,Figueroa:2020jkf,Figueroa:2021zah,Raatikainen:2023bzk,Jackson:2022unc,Jackson:2024aoo,Vennin:2024yzl}) but that are inherently captured by lattice simulations, as well as the nonlinearities arising from the relation between the inflaton field fluctuation \(\delta\phi\) and the curvature perturbation \(\zeta\) at the end of inflation, as described by the $\delta N$ formalism. 
	
	Our final results consist of three-dimensional maps of the fully nonlinear curvature perturbation at the end of inflation, from which we extract the relevant curvature perturbation statistics.
	We compare these results to those obtained using standard perturbative reasoning based on a gauge transformation. We stress that the latter approach is intrinsically perturbative, and we explicitly demonstrate its breakdown in the presence of large fluctuations. We then motivate an analytical framework that extends its regime of validity via an effective background description. This work represents a crucial milestone toward a robust calculation of SIGWs and PBHs from inflation using lattice simulations. The code used in this study is available at the following \href{https://github.com/caravangelo/inflation-easy.git}{link}, with additional details provided in the accompanying publication~\cite{Caravano:2025klk}.
	
	This paper is organized as follows: In Sec.~\ref{sec:setup}, we summarize the setup used in Paper I. In Sec.~\ref{sec:computing-zeta}, we describe our method for nonperturbatively computing $\zeta$ from the lattice simulations and compare it with other approaches. In Secs.~\ref{sec:results} and \ref{sec:discussion}, we present and discuss our results, and in Sec.~\ref{sec:conclusions}, we conclude and outline future directions. We set $M_{\textrm{Pl}}^2=1/(8 \pi G)=1$ throughout.
	
	\section{Modeling the inflationary USR phase and its self-interactions}
	\label{sec:setup}
	
	In this introductory section, we summarize the setup adopted in Paper I, which forms the basis for the present work. We consider a representative toy model of single-field inflationary scenarios featuring a transient phase of USR. The inflaton field $\phi$ is minimally coupled to gravity, and the model is fully characterized by the inflaton potential $V(\phi)$.
	
	In the framework of standard perturbation theory, the inflationary background is described by the evolution of the Hubble rate $H \equiv \dot{a}/a$, where $a$ is the scale factor during inflation and overdots denote derivatives with respect to cosmic time $t$. This evolution is encoded in the Hubble parameters
	\begin{align}
		\label{eq:HubbleParameters}
		\epsilon \equiv -\frac{\dot{H}}{H^2}\,, \qquad
		\eta \equiv -\frac{\ddot{H}}{2H\dot{H}} = \epsilon - \frac{1}{2}\frac{d\log\epsilon}{dN}\,,
	\end{align}
	where $N$ denotes the number of $e$-folds, such that $dN = Hdt$. These parameters are connected to an alternative definition of the second ``slow-roll'' parameter commonly used in the literature: $\epsilon_2 \equiv \dot \epsilon / (H \epsilon) = -2 (\eta - \epsilon)$.
	
	To capture the phenomenology of generic USR scenarios, we reverse-engineer the USR phase by parameterizing the evolution of the Hubble parameters through an analytical model for $\eta(N)$\,\cite{Byrnes:2018txb,Taoso:2021uvl,Franciolini:2022pav}. In this construction, the inflaton initially undergoes SR down its potential, eventually reaching an approximate inflection point that triggers a USR phase characterized by $\eta \gtrsim 3/2$. This is followed by a second SR phase with a small negative value of $\eta$. Throughout the evolution, the parameter $\epsilon$ remains small.
	
	This behavior is modeled using a three-stage piecewise constant function for $\eta(N)$ with smooth transitions between the regimes: an initial SR phase controlling CMB scales ($\eta_{\rm I} \ll 1$), a transient USR phase ($\eta_{\rm II} \gg 3/2$), and a final SR phase ($\eta_{\rm III} \leq 0$). From this prescription for $\eta(N)$, the corresponding inflaton potential can be reconstructed as detailed in \cite{Franciolini:2022pav}.
	
	To explore USR phases that produce different amplitudes of the curvature power spectrum and non-Gaussianities, we vary the duration of the USR phase $\Delta N\equiv N_{\rm end} - N_{\rm in}$, as well as the value of $\eta_{\rm II}$. Since perturbations grow exponentially with $N$ during USR, a longer duration leads to a larger power spectrum and stronger non-Gaussianities. The parameters used in our analysis are listed in Tab.~\ref{tab:usr_cases}; see Paper I for further details.
	
{
	\renewcommand{\arraystretch}{1.2}
	\setlength{\tabcolsep}{4pt}
	
	\begin{table}
		\caption{Parameters adopted in the various scenarios considered in this work. The value of $\Delta N$ is expressed as a function of the maximum tree-level curvature power spectrum $\mathcal{P}_\zeta^{\rm max}$.}
		\label{tab:usr_cases}
		\begin{ruledtabular}
			\begin{tabular}{lccc}
				& $\eta_{\rm II}$ & $\eta_{\rm III}$ & $\Delta N \equiv N_{\rm end}-N_{\rm in}$ \\
				\hline
				Case I (Wands duality) & 3.5 & -0.5 & $2.6 + 0.29 \log_{10}\!\mathcal{P}_\zeta^{\rm max}$ \\
				Case II (repulsive)    & 3.0 & -0.5 & $3.3 + 0.38 \log_{10}\!\mathcal{P}_\zeta^{\rm max}$ \\
				Case III (attractive)  & 4.5 & -0.5 & $1.8 + 0.19 \log_{10}\!\mathcal{P}_\zeta^{\rm max}$ \\
			\end{tabular}
		\end{ruledtabular}
	\end{table}
}
	
	Although the lattice simulations treat the dynamics of the inflaton nonperturbatively, it is instructive to make the link with perturbative approaches. For this, note that the lattice simulations are performed in the spatially flat gauge, where
	\begin{equation}
		h_{ij}(t) = a^2(t)\delta_{ij}\,, \qquad
		\Phi(t, \vec{x}) = \phi(t) + \delta\phi(t, \vec{x})
		\label{eq:flat-gauge}
	\end{equation}
	and the total scalar field has been split into its spatial average and fluctuations about it. We neglect tensor modes and we work in the decoupling limit, neglecting metric perturbations in the lapse and shift since $\epsilon$ remains much smaller than unity throughout the evolution.

	Self-interactions of $\delta \phi$ are governed by derivatives of the potential, particularly those involving time derivatives of $\eta$ during USR. The action for the fluctuations reads:
	\begin{align}
		\label{eq:pert_act}
		S = \int d\tau\, d^3x\, a^2 \left[
		\frac{1}{2} \left(\partial_\tau \delta\phi\right)^2 
		- \frac{1}{2} \left(\partial_i \delta\phi\right)^2 
		- a^2 \sum_{n \geq 2} \frac{V_n \delta\phi^n}{n!}
		\right],
	\end{align}
	where $V_n \equiv \d^n V / \d \phi^n$. Neglecting $\epsilon$-suppressed terms, the leading orders of the potential derivatives are given by (see e.g.~\cite{Franciolini:2023lgy,Ballesteros:2024zdp}):
	\begin{subequations}
		\begin{align}
			\label{eq:couplings}
			a^2 V_2 &= -(aH)^2 (\nu^2 - 9/4)\,, \\
			a^2 V_3 &= -\textrm{sign}(\dot\phi) \frac{aH (\nu^2)'}{\sqrt{2\epsilon}}\,, 
			\label{eq:cubiccoupling} \\
			a^2 V_4 &= -\frac{1}{2\epsilon} \left[ (\nu^2)'' - aH (\nu^2)'(1 - \eta) \right]\,,
		\end{align}
	\end{subequations}
	where primes denote derivatives with respect to conformal time, and we define
	\begin{align}
		\label{eq:nu2}
		\nu^2 \equiv \frac{9}{4} - \left[ \eta(3 - \eta) + \frac{\eta'}{aH} \right].
	\end{align}
	Looking at Eq.~\eqref{eq:nu2}, it is evident that for constant $\eta$, the value of $\nu^2$ remains invariant under the transformation $\eta \to 3-\eta$, which goes under the name of Wands duality \cite{Wands:1998yp}.
	
	Depending on the value of $\eta$ during USR, we consider three cases as outlined in Tab.~\ref{tab:usr_cases}:
\begin{itemize}[noitemsep,leftmargin=*]
	\item Case I: \textbf{Wands duality}. In this scenario, $\nu^2$ remains approximately constant following the onset of USR, suppressing inflaton self-interactions and yielding a nearly free theory for $\delta\phi$ fluctuations.
	
	\item Case II: \textbf{Repulsive}. Here, the time evolution of $\nu^2$ is significant, increasing around the end of the USR phase (near $N-N_{\rm in} \sim 2$). This leads to a positive cubic coupling $V_3$ in Eq.~\eqref{eq:cubiccoupling}, associated with repulsive interactions. As a result, the inflaton perturbations exhibit non-Gaussianities that suppress large positive field excursions.
	
	\item Case III: \textbf{Attractive}. In contrast, this case features a decreasing $\nu^2$ near the USR-to-SR transition, yielding a negative cubic coupling indicative of attractive behavior. Consequently, the inflaton perturbations display non-Gaussianities that enhance large positive field excursions.
\end{itemize}

	The potentials are monotonous in case II, but they feature a local maximum and a local minimum in cases I and III. In these cases, some parts of the universe can get trapped into the local minimum, eventually forming PBHs after inflation.
	This, together with the properties of the super-Hubble  inflaton perturbations at the end of the USR, relevant to build observable late-time universe predictions, are carefully described in Paper I. We now move our attention to the relationship between these perturbations and the comoving curvature perturbations $\zeta$ describing scalar fluctuations at the end of inflation.

	\section{Computing the curvature perturbation}
	\label{sec:computing-zeta}
	
	The curvature perturbation $\zeta$ is defined by choosing slices of uniform field values to coincide with spatial hypersurfaces of constant time coordinate. In this comoving gauge, the inflaton is thus spatially uniform by definition. Neglecting vector and tensor degrees of freedom for our purposes, one can further adapt spatial coordinates such that the spatial part of the metric is diagonal, with $h_{ij}=\tilde{a}^2(\tc,\vec{x}_\textrm{c})\delta_{ij}$, and where $\textrm{c}$ stands for comoving (in a similar context, see, e.g.~\cite{Ballesteros:2024zdp} for considering these degrees of freedom). The comoving curvature perturbation is then defined such that $\tilde{a}(\tc,\vec{x}_\textrm{c})=a_\textrm{c}(\tc) e^{\zeta(\tc,\vec{x}_\textrm{c})}$, where the a priori arbitrary splitting of the local scale factor $\tilde{a}(\tc,\vec{x}_\textrm{c})$ into a homogeneous component and a spatially varying part is made by requiring that the spatial average of $\zeta$ vanishes.
	
	We aim to determine $\zeta$ at the end of inflation, or more precisely,  when the inflaton follows the slow-roll attractor and has reached an adiabatic limit, ensuring that $\zeta$ has become constant and provides a good variable to characterize primordial fluctuations left over from inflation. Note that on such attractors, the field velocity is uniquely defined as a function of the field value, so that comoving slices of uniform inflaton coincide with slices of uniform energy density. The $\delta N$ formalism \cite{Starobinsky:1985ibc,PhysRevD.42.3936,Sasaki:1998ug,Wands:2000dp,Lyth:2004gb,Lyth:2005fi} then provides us with a straightforward method to determine $\zeta$, taking as an input the three-dimensional real-space maps of the inflaton field and its derivative computed with the lattice simulations.

	\subsection{Nonperturbative computation with $\delta N$ formalism}
	
	At the end of the lattice simulations, performed in the spatially flat gauge, each of the \(N^3_{\rm grid}\) cells has become larger than the comoving Hubble radius, and effectively evolves as a separate universe. The curvature perturbation $\zeta$ precisely measures the relative amount of \textit{e}-folds of expansion between these separate universes, from this initial flat hypersurface to the final hypersurface of constant energy density, or equivalently of constant inflaton.\footnote{We recall that the local number of \textit{e}-folds between any two spatially flat hypersurfaces is actually uniform so that the initial flat hypersurface of the $\delta N$ formalism can be chosen arbitrarily.}
	
	In this separate-universe picture, we can evolve the system using the following equation  
	\begin{align}\label{eq:EoM}
		\frac{d^2\phi}{dN^2} + \left[3 - \frac{1}{2}\left(\frac{d\phi}{dN}\right)^2\right]
		\left[\frac{d\phi}{dN} + \frac{V_{,\phi}(\phi)}{V(\phi)}
		\right] = 0\,,
	\end{align}
	which is the Klein-Gordon equation for the inflaton, neglecting gradients and using the first Friedmann equation to express the Hubble parameter. To obtain \(\zeta\) in the late-time limit, we numerically integrate this equation—using initial conditions for the field and its derivative from the lattice simulations that capture the USR dynamics—up to a fixed field value on the slow-roll part of the potential. Although the final result is independent of the precise value chosen, this field value is typically taken to be the one that minimizes the energy at the end of the lattice simulation, as explained in~\cite{Caravano:2025klk}. We record the number of \(e\)-folds required to reach this point, constructing a three-dimensional map of \(N(\vec{x})\). The comoving curvature perturbation then follows as  
	\begin{equation}
		\zeta(\vec{x}) = N(\vec{x}) - \langle N(\vec{x}) \rangle,  
		\label{eq:zeta-delta-N}
	\end{equation}  
	where \(\langle \cdot \rangle\) denotes a lattice spatial average.

	So far, our discussion has been generic. However, an important subtlety in our context is that $\zeta$ is not necessarily defined everywhere. This is simple to understand: for some of the models we consider, some parts of the universe get trapped into a local minimum of the potential. Inflation never ends in these trapped patches, so the local number of $e$-folds to finish inflation is simply not defined. Another perspective on this is that the inflaton is used as a clock field in the comoving gauge, as slices of constant field values are used as slices of constant time variable. However, for trapped patches, the value of the inflaton decreases and then increases as the inflaton rolls back towards the local minimum. Hence, the 
	comoving gauge cannot be globally defined in these models, as the clock stops ticking at the turnaround point.  An interesting generic consequence of this phenomenon is that $\zeta$ grows without bound as one considers patches that are close to being trapped, something we will discuss in detail below.

	The procedure described in this section—and used to compute $\zeta$ from the lattice in this work—is intrinsically nonperturbative: the simulations solve the full inflaton dynamics from the Bunch-Davies vacuum to the super-Hubble regime, while the $\delta N$ formalism enables a fully nonperturbative extraction of the curvature perturbation at the end of inflation. Nevertheless, it is instructive to compare our nonperturbative approach with a more conventional perturbative argument based on a change of gauge, as we do in the next section, to better understand the breakdown of perturbation theory. While we applied this method to USR inflation in this paper, it is general and can be applied to any early-universe scenario, including those studied in \cite{Caravano:2021pgc,Caravano:2022yyv,Caravano:2024tlp,Caravano:2022epk,Caravano:2024xsb}. 
	
	{Before moving on, let us briefly comment on the role of metric perturbations in the lattice computations of \cite{Caravano:2024moy}. In our simulations we work in the decoupling limit, where the metric is taken to be of the FLRW form with negligible perturbations. This implies neglecting, for example, local variations in the expansion rate. Such an approximation is well justified—even for very large field excursions—since the energy density is potential dominated and the potential itself is extremely flat. Nevertheless, the approximation can in principle break down, and it would be interesting to go beyond it, either by including perturbative corrections as in \cite{Caravano:2024xsb,Jamieson:2025ngu} or by resorting to numerical relativity. In practice, however, the results are expected to be indistinguishable, as recently demonstrated in the case of a resonant potential: despite highly inhomogeneous dynamics, the lattice simulations of \cite{Caravano:2024tlp} show excellent agreement with the corresponding numerical relativity analysis of \cite{Launay:2025kef}.\footnote{See also \cite{Creminelli:2024cge} for a quantitative study of the validity of the decoupling limit in the far tail.}}

	\subsection{Change of gauge}
	\label{sec:change-gauge}
	
	A standard approach in perturbation theory is to interpret the relationship between the curvature perturbation and inflaton inhomogeneities as a change of gauge. In this section, we review this argument and stress its perturbative nature.
	
	We start with the inhomogeneous scalar field $\Phi(t, \vec{x})$ in the spatially flat gauge, of spacetime coordinates $x^\mu=(t,\vec{x})$, and our goal is to identify the coordinate system $x_\textrm{c}^\mu=(\tc,\vec{x}_\textrm{c})$ in which the inflaton field appears homogeneous, and the spatial part of the metric takes the form $h_{ij}=a_\textrm{c}(\tc) e^{\zeta(\tc,\vec{x}_\textrm{c})}\delta_{ij}$.
	The link between the two systems of coordinates is very complicated in general, but it drastically simplifies on super-Hubble scales, where gradients are negligible and all choices of spatial coordinates become equivalent, see e.g.~\cite{Maldacena:2002vr,Behbahani:2011it}.
	One then has 
	$
	\zeta=\ln \left (
	{a_\textrm{c}(\tc+\delta t)}/{a_\textrm{c}(\tc)}
	\right ),$
	where $\delta t=t-\tc$ is the time shift between the flat and the comoving slicings, such that $\Phi(t, \vec{x})=\phi_\textrm{c}(t_c)$. The procedure at this stage is purely geometrical and entirely nonperturbative, but it is not operational, as the one one-to-one identification $\phi_\textrm{c}(t_c)$ between field values and the time coordinate in the comoving gauge can be a priori chosen in an arbitrary (monotonous) manner, and $a_\textrm{c}(t_c)$ is unknown. In particular, one can choose the function $\phi_\textrm{c}(t_c)$ to have the same functional dependence as $\phi(t)=\langle \Phi(t, \vec{x}) \rangle$, the spatial average of the inflaton on flat slices, see Eq.~\eqref{eq:flat-gauge}. However, there is no justification for identifying the two functions $a(t)$ and $a_\textrm{c}(\tc)$—that is, the scale factor as a function of time in the flat gauge and the average scale factor on comoving slices, defined by $a_\textrm{c}(\tc) = \exp \langle \ln( \tilde{a}(\tc,\vec{x}_\textrm{c}) ) \rangle$. In this case, the relation above does not allow one to compute $\zeta$ in any meaningful way.
	
	An operational link between the two systems of coordinates can only be made in perturbation theory: one then assumes the same functional forms for $\phi(t)=\langle \Phi(t, \vec{x}) \rangle$ and $\phi_\textrm{c}(t_c)$, as well as for $a(t)$ and $a_\textrm{c}(\tc)$. However, one cannot know a priori the regime of validity of that perturbative approach, i.e.~one is not guaranteed that the resulting computation will correctly describe $\zeta$ for large inhomogeneities. With that perturbative ansatz at hand, one can further specify the time coordinate by imposing conditions on the spacetime metric $ds^2=-{\cal N}^2 d\tc^2+a^2_\textrm{c}(\tc) e^{2\zeta(\tc,\vec{x}_\textrm{c})}\delta_{ij}(dx^i+N^i d \tc)(dx^j+N^j d \tc)$. The choice of cosmic time corresponds to imposing ${\cal N}^2=1+\cal{O}(\zeta)$, in which case the constraint equation for the lapse ${\cal N}$, at zeroth order in an expansion in $\zeta$, gives the conventional Friedmann equation $3(\dot a/a)^2=\frac12 \dot{\phi}^2+V(\phi)$. Alternatively, one can choose the number of $e$-folds such that $\dot a=a$, in which case the constraint equation for the lapse, at zeroth order in an expansion in $\zeta$, gives ${\cal N}^2=\frac{3-\frac12 \dot{\phi}^2}{V(\phi)}(1+\cal{O}(\zeta))$.
	This choice is particularly convenient and we adopt it in the following, conventionally using the symbol $N$ for this time coordinate. One then obtains 
	\begin{equation}
		\begin{aligned}
			\zetapt(N,\vec{x})=\delta N \,, \quad & \textrm{such that}\quad   \Phi(N,\vec{x})=\phi(N-\delta N)\\
			&\textrm{with} \qquad  \quad \,\phi(N) \equiv \langle \Phi(N,\vec{x}) \rangle\,.
		\end{aligned}
		\label{eq:change-gauge}
	\end{equation}
	This equation expresses the super-Hubble comoving curvature perturbation as the shift in the number of $e$-folds required to go from the spatially flat gauge to the comoving one. The reason why we explained this simple-looking equation in such detail is now hopefully clear: we wanted to stress the often implicit assumptions behind it, and in particular that, maybe despite appearances, this relationship is 
	inevitably perturbative. It is for this reason that we denote $\zetapt$, for perturbation theory, this perturbative computation of the curvature perturbation.

	In Sec.~\ref{sec:quantitative-break-down-PT}, we will numerically solve Eq.~\eqref{eq:change-gauge} and show explicitly its deviation from the nonperturbative computation \eqref{eq:zeta-delta-N} using the $\delta N$ formalism. 
	However, the differences between the two are readily apparent and worth stressing. First, the reasoning based on a change of gauge relates two slicings at the same spacetime point. Hence it is local in that sense, although to make it operative we need to make a spatial average on the flat slicings. By contrast, the $\delta N$ formalism is global, in the sense that it relates different slicings at different spacetime points, the ``initial'' flat hypersurface and the ``final'' hypersurface of constant field value where the inflaton has reached an attractor.
	Second, $\zetapt$ deduced from the change of gauge relation~\eqref{eq:change-gauge} only depends on the local field value, whereas in the full nonperturbative $\delta N$ formalism, the number of $e$-folds left to reach the final constant $\phi$ hypersurface
	a priori depends on the field value \textit{and} its time derivative. Equivalently, Eq.~\eqref{eq:change-gauge} can correctly predict $\zeta$ only if all parts of the universe follow exactly the same history \textit{at all times}---the one of the ``background'' $\phi(N)$---which is never the case, although it provides an excellent approximation for typical regions that probe a limited range of the dynamics. By contrast, the $\delta N$ formalism does not make such a hypothesis but precisely, it determines how the cosmological history depends on initial phase-space configurations, until all points have reached an attractor and the relative amount of expansion $\zeta$ has become constant.

	\subsection{Effective background description}
	\label{sec:effective-background}
	
	We have explained that conventional reasonings based on a change of gauge are necessarily limited. Nevertheless, in this section, we solve Eq.~\eqref{eq:change-gauge} using an effective background dynamics characterized by a constant $\eta$ \textit{at all times}. This simple model yields a simple analytical expression that is expected to accurately predict $\zeta$ for the bulk of lattice points that, by the end of the simulation, have reached the second slow-roll attractor (characterized by constant $\eta=\eta_{\textrm{III}} = -0.5$). Moreover, as we will see in Sec.~\ref{sec:results} and discuss in Sec.~\ref{sec:discussion}, this description, which locally reduces to the background but that strongly differs from it in the past, actually provides insight into the physics of large fluctuations.

	To avoid confusion with the true background $\phi(N)$, we will use the symbol $\bar{\phi}(N)$ for the effective model. Denoting by primes derivatives with respect to the number of $e$-folds and recalling that $\epsilon = \frac12 \bar{\phi}^{\prime 2}$ as well as Eq.~\eqref{eq:HubbleParameters}, one has in general \(\bar\phi^{\prime\prime} = (\epsilon - \eta) \bar\phi^{\prime}\). The effective description of interest corresponds to a fiducial background dynamics where $\eta$ is constant, negative, at all times, and $\epsilon \ll \eta$ is negligible, in which case one finds upon integration 
	\begin{align}
		\label{eq:fiducial-background}
		\begin{split}
			& \bar\phi^\prime(N) = \bar\phi^\prime(N_0) e^{-\eta(N - N_0)}, \\
			& \bar\phi(N) = \bar\phi(N_0) - \frac{\bar\phi^\prime(N_0)}{\eta} \left( e^{-\eta(N - N_0)} - 1 \right),
		\end{split}
	\end{align}
	where the time \(N_0\) is arbitrary. To better grasp the physics at play, notice that this trajectory corresponds to the dynamics of a scalar field evolving on a potential with the shape of an inverted parabola and under specific boundary conditions. Let us assume indeed a potential of the form $V(\phi) =\bar{V}-\frac{1}{2} m^2 \phi^2$, taking the maximum of the potential at $\phi=0$ without loss of generality. From the Klein-Gordon equation $\bar{\phi}^{\prime\prime}+3 \bar{\phi}^\prime(1-\epsilon)+V'/H^2$, and self-consistently neglecting $\epsilon$ and writing $3 H^2=\bar{V}$, one sees that the two descriptions agree for $m^2=-\frac13 \eta(3-\eta) \bar{V}$, i.e.
	\begin{equation}
		\label{eq:parabola}
		V(\phi) =\bar{V} \left(1+\frac{1}{6} \eta(3-\eta)\phi^2 \right)\,.
	\end{equation} 
	More specifically, Eq.~\eqref{eq:fiducial-background} corresponds to the solution with $\bar \phi' \to 0$ as $\bar \phi \to 0$ and $N \to -\infty$, i.e.~to a scalar field released asymptotically at the top of the inverted quadratic potential \eqref{eq:parabola} with vanishing velocity in the infinite past. Notice also that the potential is invariant under the Wands duality \(\eta \rightarrow 3 - \eta\) something to which we will come back in Sec.~\ref{sec:discussion}.
	
	Now, it is straightforward to solve Eq.~\eqref{eq:change-gauge} for the fiducial background \eqref{eq:fiducial-background}, and to interpret its result. More precisely, let us write the total scalar field value as $\Phi(N,\vec{x})=\bar{\phi}(N)-\bar{\phi}^\prime(N) \zeta_{\rm lin}$. Namely, we parametrize it in terms of the ``linear'' comoving curvature perturbation \(\zeta_{\rm lin} \equiv -\delta\phi / \bar\phi^\prime\) that one would find by a change of gauge at first order in perturbation theory around that background dynamics, hence the subscript ``${\rm lin}$''. Substituting this, together with Eq.~\eqref{eq:fiducial-background}, into Eq.~\eqref{eq:change-gauge}, we obtain:
	\begin{equation}
		\label{eq:log}
		\zeta(\vec{x}) = \frac{1}{\eta} \log(1 + \eta \, \zeta_{\rm lin}(\vec{x}))\,,
	\end{equation}
	where one sees that the relationship between $\zeta_{\rm lin}$ and $\zeta$ does not depend on the specific time $N_*$ at which the gauge transformation is performed.

	\begin{figure}
		\centering
		\includegraphics[width=1 \linewidth]{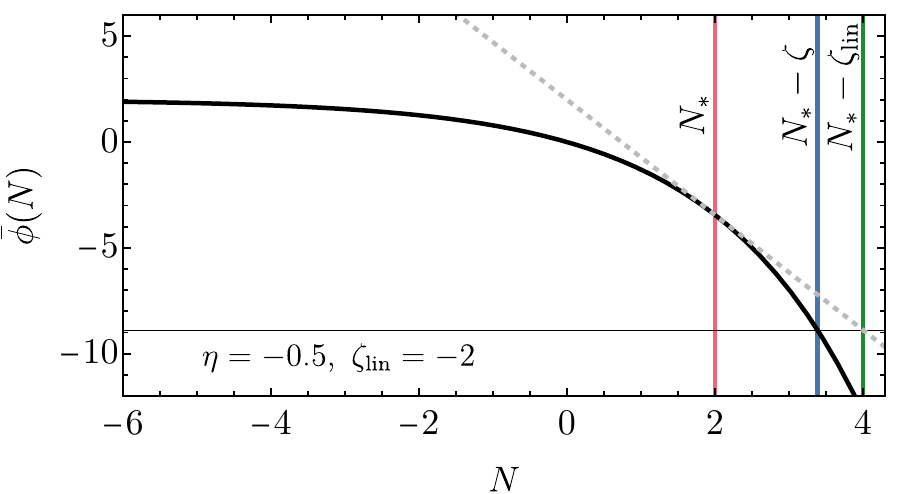}
		\includegraphics[width=1 \linewidth]{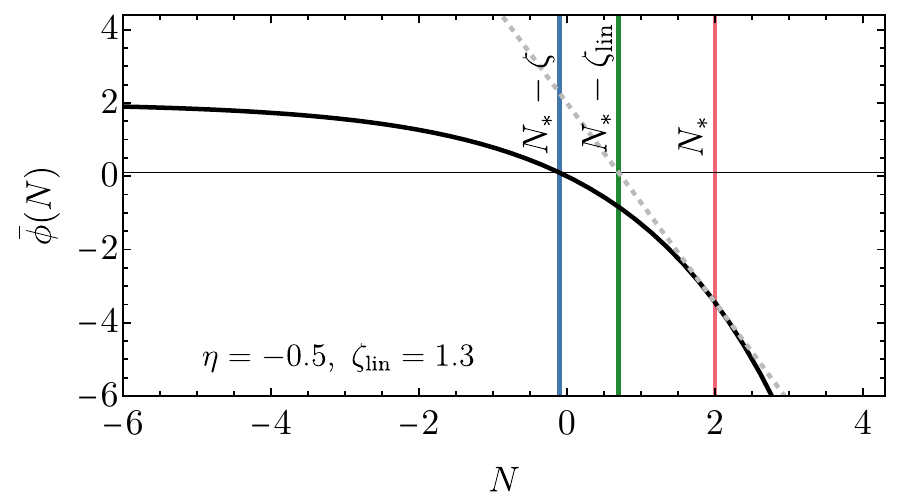}
		\includegraphics[width=1 \linewidth]{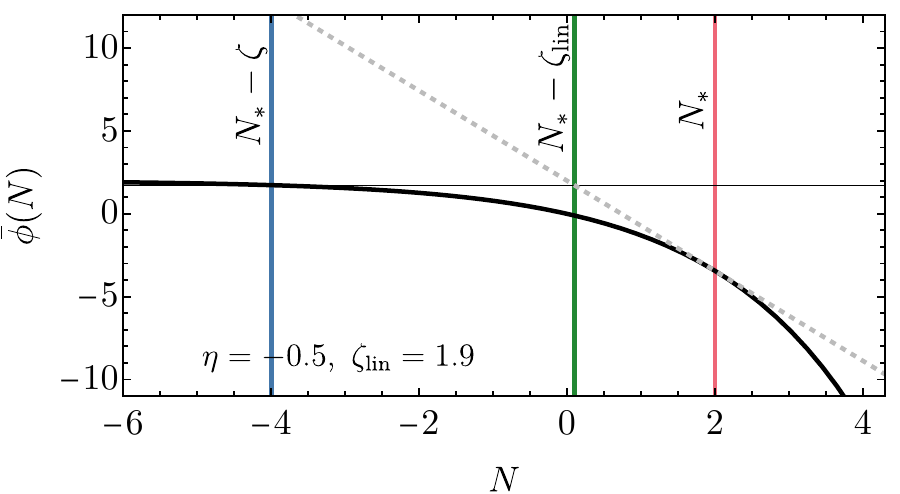}
		\caption{Graphical representation of the change of gauge relationship \eqref{eq:change-gauge} for the fiducial background \eqref{eq:fiducial-background} for $\eta=-0.5$. The change of gauge is performed at $N_*=2$ and the three plots correspond to different values of the scalar field in the flat gauge. See the main text for more details.}
		\label{fig:figure-zeta-log}
	\end{figure}

	We now present a simple geometrical interpretation of this relation that clarifies the physical origin of its logarithmic divergence. It is illustrated in Fig.~\ref{fig:figure-zeta-log} for $N_*=2$ and arbitrary values of $\bar\phi(N_0)$ and $\bar\phi'(N_0)$.
	For a given value of $\Phi(N_*,\vec{x})$, represented in each case by the faint horizontal black curve, its intersection with the dark black curve $\bar{\phi}(N)$ is at the location $N_*-\zeta$, marked by the blue line. By contrast, its intersection with the tangent of $\bar{\phi}(N)$ at $N_*$ is at the location $N_*-\zeta_{\textrm{lin}}$, marked by the green line. For small fluctuations, the curve can be approximated by its tangent and $\zeta \approx \zeta_{\textrm{lin}}$, but the two strongly differ for large fluctuations. The logarithmic divergence of \eqref{eq:log} near \(\zeta_{\rm lin} = -1/\eta\) is readily apparent in this figure. The background dynamics with constant $\eta$ corresponds to a vanishing velocity in the far past, near the top of the inverted parabola. For field values that approach this configuration (see the bottom panel in Fig.~\ref{fig:figure-zeta-log}), $\zeta$, the number of $e$-folds it takes to reach $\phi(N_*)$ along that curve, corresponding to the interval between the blue and the red lines, grows without bound. On the other hand, no singularity is reached for negative values of $
	\zeta$ (or $\zeta_{\rm lin}$) as any arbitrary large (negative) field value can always be attained with a sufficiently large $\zeta_{\rm lin}$, see the top panel of Fig.~\ref{fig:figure-zeta-log}.
	
	\begin{figure*}
		\centering
		\includegraphics[ width=0.367 \linewidth]{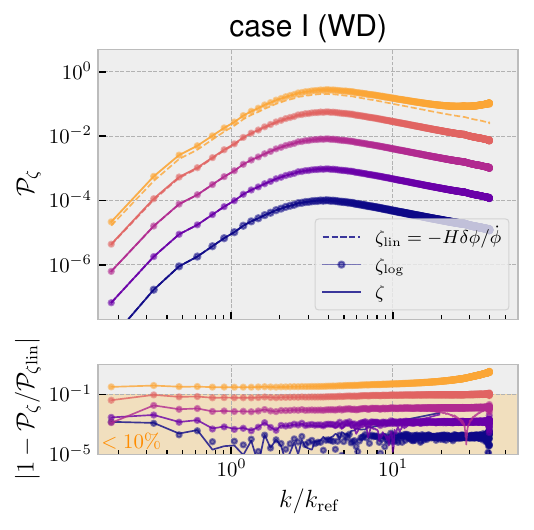}
		\includegraphics[width=0.31 \linewidth]{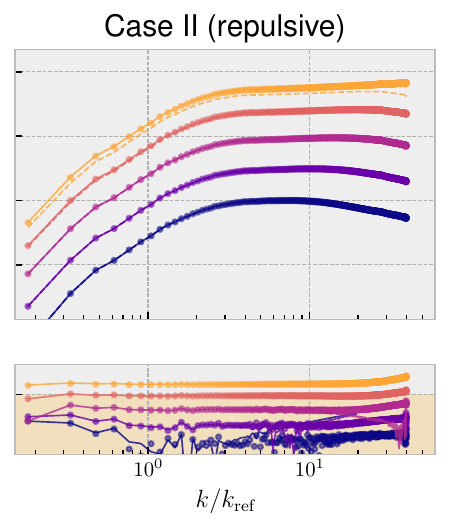}
		\includegraphics[width=0.31 \linewidth]{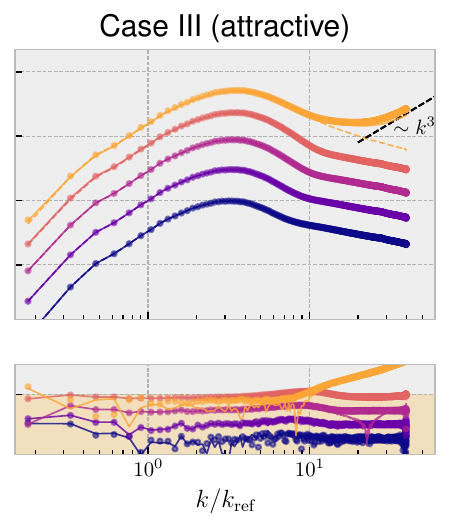}
		\caption{Power spectrum of the comoving curvature perturbation $\zeta$ from the lattice simulation. Full lines represent the nonperturbative lattice results obtained with the $\delta N$ approach, dashed lines the linearized perturbation $\zeta_{\rm lin}=-H\delta\phi/\dot\phi$ and the dots show results from the logarithmic relation in Eq.~\eqref{eq:log} applied to $\zeta_{\rm lin}$. Different colors correspond to different lattice simulations presented in Paper \cite{Caravano:2024moy}, ranging from ${\cal P}_{\zeta,\rm tree}^{\rm  max} =10^{-4}$ to ${\cal P}^{\rm  max}_{\zeta,\rm tree} =1$, where $\cal P_{\zeta,\rm tree}$ is the tree-level power spectrum.}
		\label{fig:PS}
	\end{figure*}

	\section{Results}
	\label{sec:results}

	\begin{figure*}
		\includegraphics[width=18cm]{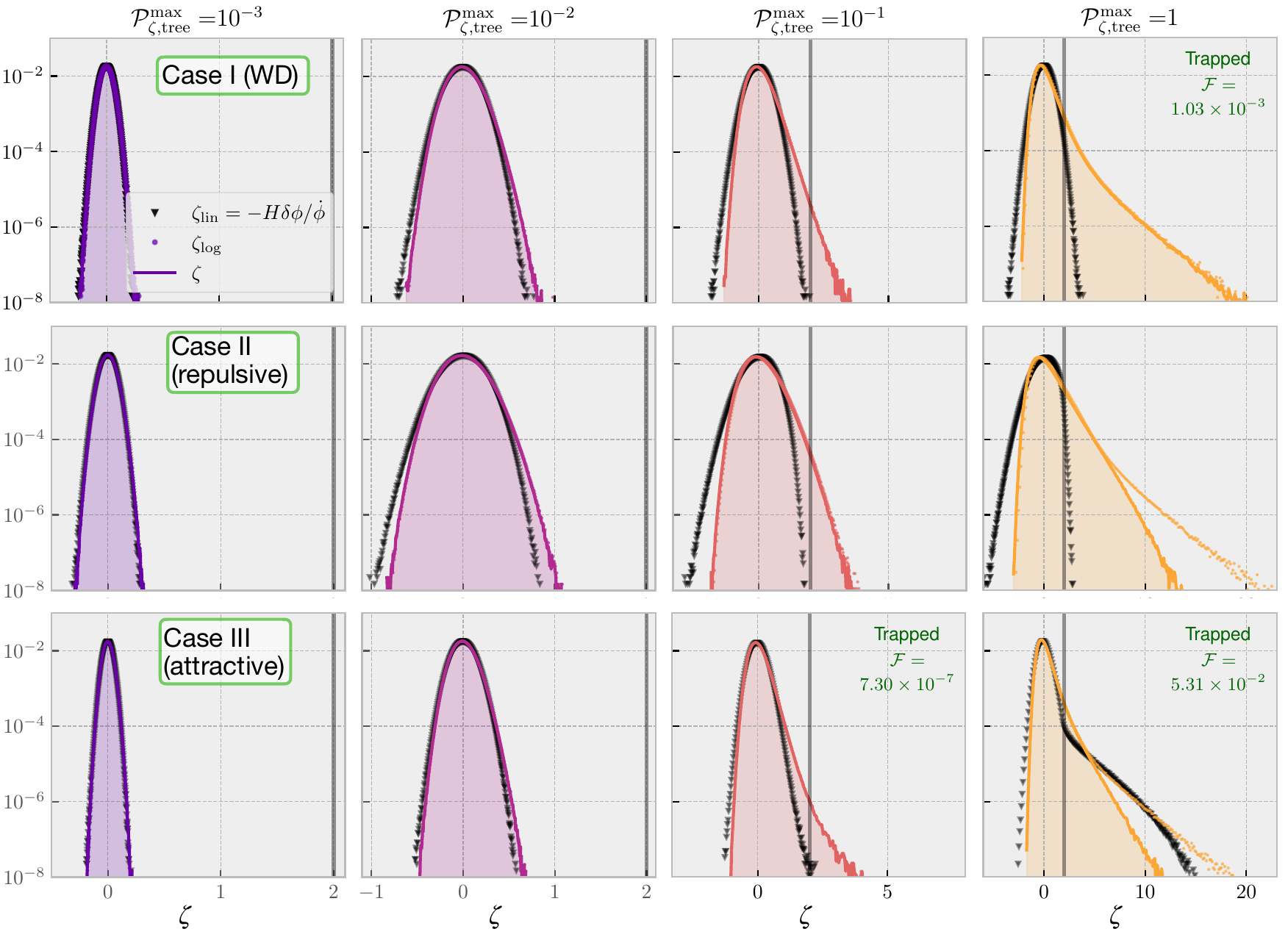}
		\caption{1-point probability density function (PDF) of the comoving curvature perturbation $\zeta$. Different columns correspond to different $\mathcal{P}^{\rm max}_{\zeta,\rm tree}$ ranging from $10^{-3}$ to $1$, while the rows refer to the three different cases. Full lines represent the nonperturbative lattice results obtained with the $\delta N$ approach, black triangles correspond to the linearized perturbation $\zeta_{\rm lin}=-H\delta\phi/\dot\phi$, and the dots show results from the logarithmic relation in Eq.~\eqref{eq:log} applied to $\zeta_{\rm lin}$. Vertical gray lines indicate the value $\zeta=2$, where Eq.~\eqref{eq:log} diverges if applied to $\zeta_{\rm lin}$.}
		\label{fig:PDF}
	\end{figure*}
	
	In this section, we present results for the curvature perturbation from all 15 simulations discussed in Paper I. Each of the three cases described in Sec.~\ref{sec:setup} includes five simulations, with tree-level power spectra spanning \(\mathcal{P}_{\zeta,\rm tree}^{\rm max} = 10^{-4}\) to 1 (by this, we refer to the power spectrum computed at linear order in standard perturbation theory, without backreaction taken into account). We note that, in addition to its intrinsic interest, exploring theories with ${\cal P}_\zeta \approx 1$ makes it easier to study phenomena associated with very large fluctuations, which also occur at smaller values of the power spectrum but are much rarer and therefore harder to access statistically.
	
	The simulations yield three-dimensional maps of the inflaton field and its velocity on a $512^3$ grid, with comoving box size $L$ chosen such that the simulation captures the following range:
	\begin{align}
		\label{eq:res}
		0.22 \,k_{\rm ref} \leq k \leq 45\, k_{\rm ref},
	\end{align}
	where $k_{\rm ref}=a_{\rm ref}H_{\rm ref}$ is the mode that exits the Hubble radius at the beginning of the USR phase, defined as $N=N_{\rm ref} = 0$ (see \cite{Caravano:2024moy} for details).  The starting point of the calculation presented in this work is the phase-space configuration, i.e.~the field and its time derivative, at the final simulation time, $N = 5$, at which point the smallest scale simulated is super-Hubble, and gradient effects are negligible. From this configuration, we compute $\zeta$ using the procedure described in the previous section.

	The fully nonperturbative $\delta N$ results will be compared with both the linear curvature perturbation, \(\zeta_{\rm lin}(\delta\phi) = -H\delta\phi / \dot\phi\), and the logarithmic relation in Eq.~\eqref{eq:log}, which we refer to as \(\zeta_{\rm log}(\delta\phi)\). Note that in the latter two cases, \(\delta\phi\) and \(\dot\phi\) are not computed from linear perturbation theory, but are obtained directly from the lattice simulation, which incorporates the nonlinear properties of \(\phi\) and the backreaction effects described in Paper I. In other words, this work focuses on the nonlinearity arising from the relation between \(\delta\phi\) and \(\zeta\),\footnote{We often use these terms for simplicity, but we stress that the results crucially depend on the full phase space information \((\delta\phi,\delta\phi')\).} rather than that induced by the nonlinear field evolution, which was explored in detail in Paper I.

	\subsection{Power spectrum}
	As a first summary statistic, we analyze the impact of the nonlinear relation on the scalar power spectrum.  
	From Fig.~\ref{fig:PS}, we see that, except for the most extreme case with \(\mathcal{P}_{\zeta,\rm tree}^{\rm max} = 1\), the nonlinear mapping between \(\phi\) and \(\zeta\) introduces only minor corrections to the power spectrum, always remaining below \(\sim10\%\). These corrections are significantly smaller than those induced by the backreaction on \(\dot\phi\), which we studied in Paper I \cite{Caravano:2024moy}. We also compare the lattice result obtained using the \(\delta N\) approach with the logarithmic formula \(\zeta_{\rm log}\) in \eqref{eq:log} based on the effective background with constant \(\eta = \eta_{\textrm{III}}\) at all times. In all cases, this effective description accurately reproduces the nonlinear mapping at the level of the power spectrum.

	The nonlinear relation substantially affects the power spectrum only in the most extreme case, \(\mathcal{P}_{\zeta,\rm tree}^{\rm max} = 1\), where small-scale deviations from the linear result become significant. These deviations are driven by nonperturbative effects, closely related to the local trapping of the inflaton field, a phenomenon discussed in Paper I and further below. 
	To compute these power spectra, we masked out trapped regions setting them to $\zeta=0$, to ensure not to introduce nonphysical artifacts. The small-scale deviation is then interpreted as a nonperturbative effect arising from extreme field values: regions far from the mean, but not far enough to become trapped in the USR feature, spend considerably more time at the top of the USR phase than the background. In case III, which is the attractive case and where the trapping effect is the most significant, we find that the small-scale enhancement of the power spectrum follows a Poisson noise scaling \(\sim k^3\), suggesting that the deviation is driven by small, localized regions with exceptionally large values of \(\zeta\). 
	In other words, the perturbation amplitude is mostly influenced by its location in space, i.e. whether or not the lattice point is close to a trapped patch, rather than the field statistics. Therefore, it inherits a characteristic shot-noise power spectrum.
	This interpretation agrees with the fact that the trapping is a small-scale phenomenon, which, as we discuss in more detail below, is confirmed by the real-space snapshots of \(\zeta\) in Fig.~\ref{fig:snapshots}.

	\subsection{1-point PDF and real-space maps of $\zeta$}
	As we have seen above, at the level of the power spectrum, the nonlinear relation does not induce significant changes beyond the most extreme case, \(\mathcal{P}_{\zeta,\rm tree}^{\rm max} = 1\). However, the nonlinear mapping between \(\zeta\) and \(\phi\) strongly affects non-Gaussianity. {This is evident in Fig.~\ref{fig:PDF}, which shows the one-point probability density function (PDF) of $\zeta$, obtained from the simulations as normalized histograms of $\zeta$ across the 3D box. Each histogram corresponds to a single simulation run.\footnote{More details on how these PDFs are computed can be found in \cite{Caravano:2025klk}.}} The nonperturbative results (solid lines) are compared to the linear relation (black triangles) and to the logarithmic relation \eqref{eq:log} from the effective background analysis. The latter is in excellent agreement with the exact results, except for very large fluctuations encountered for \(\mathcal{P}_{\zeta,\rm tree}^{\rm max} = 1\) in the right column, in Cases~II and III; in Case~I, by contrast, it shows excellent agreement even for arbitrarily large fluctuations. We will return to this point in Sec.~\ref{sec:regime-validity-fiducial-background}. For now, we note that in both Cases~II and III, the logarithmic relation substantially overestimates the tail of the probability distribution.

	From these plots, we learn that the fully nonperturbative relation always induces positive non-Gaussianity, meaning skewing the distribution toward larger values, which may be attributed to the gravitational interaction being attractive. Notably, this occurs even in case II, where the field self-interaction is repulsive and, in the absence of the nonlinear relation, the field distribution is negatively skewed, as shown by the black PDF in Fig.~\ref{fig:PDF} and as discussed in Paper I.

	\begin{figure*}
		\includegraphics[width=18cm]{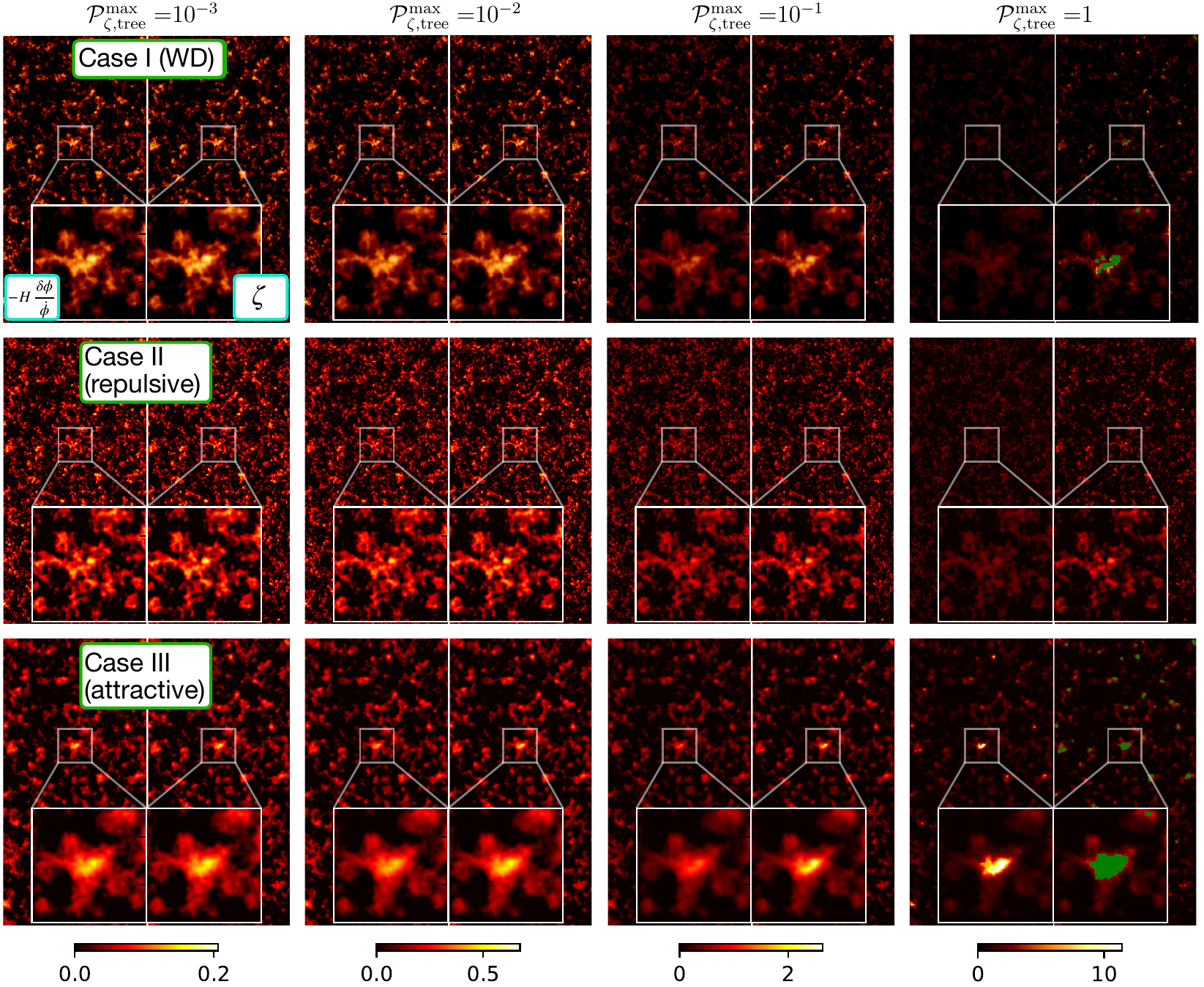}
		\caption{Snapshots from the lattice simulations. For each panel, the left sub-panel shows the curvature perturbation obtained from the linear change of gauge $\zeta_{\textrm{lin}}=-H {\delta\phi}/{\dot\phi}$, while the right sub-panel shows the nonperturbative curvature perturbation $\zeta$. Different lines show the three different cases, while the columns are for different values of the tree-level power spectrum. The green regions indicate the trapped patches. Given the wide range of values of $\zeta$, to ease visualization, we only show positive values of $\zeta$ and set negative values to zero in this figure. In Fig. \ref{fig:snapshots_negative}, we show the negative values for completeness. In Fig.~\ref{fig:3d_snapshots}, we show a 3D visualization of these snapshots for $\mathcal{P}_{\zeta,\rm tree}^{\rm max} = 1$ (right column of this figure). }
		\label{fig:snapshots}
	\end{figure*}

	\begin{figure*}
		\includegraphics[width=18cm]{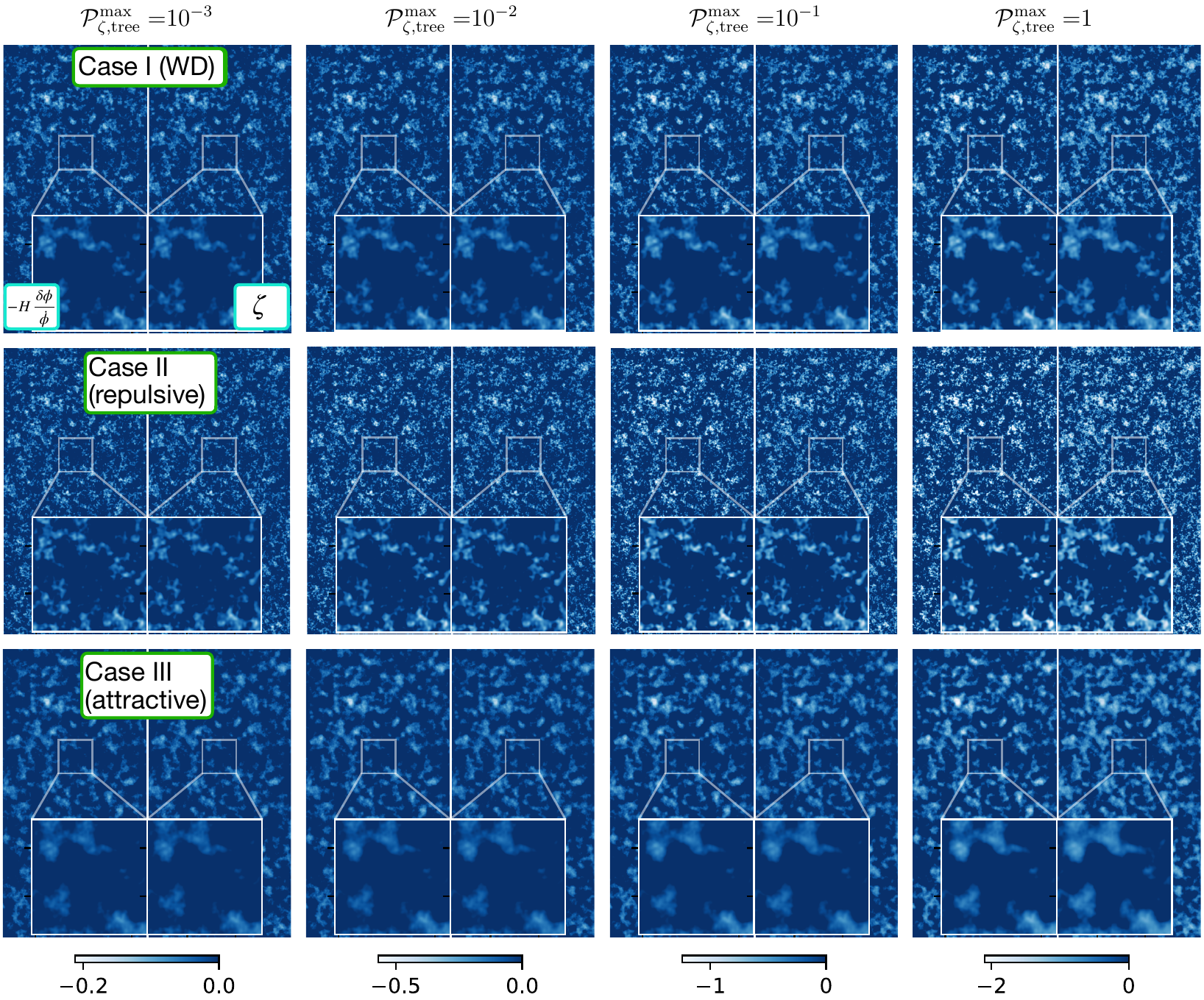}
		\caption{Same as Fig. \ref{fig:snapshots}, but for negative values of $\zeta$. In Fig.~\ref{fig:3d_snapshots}, we show a 3D visualization of these snapshots for the case $\mathcal{P}_{\zeta,\rm tree}^{\rm max} = 1$ (right column of this figure). }
		\label{fig:snapshots_negative}
	\end{figure*}
	
	\begin{figure*}
		\includegraphics[width=18cm]{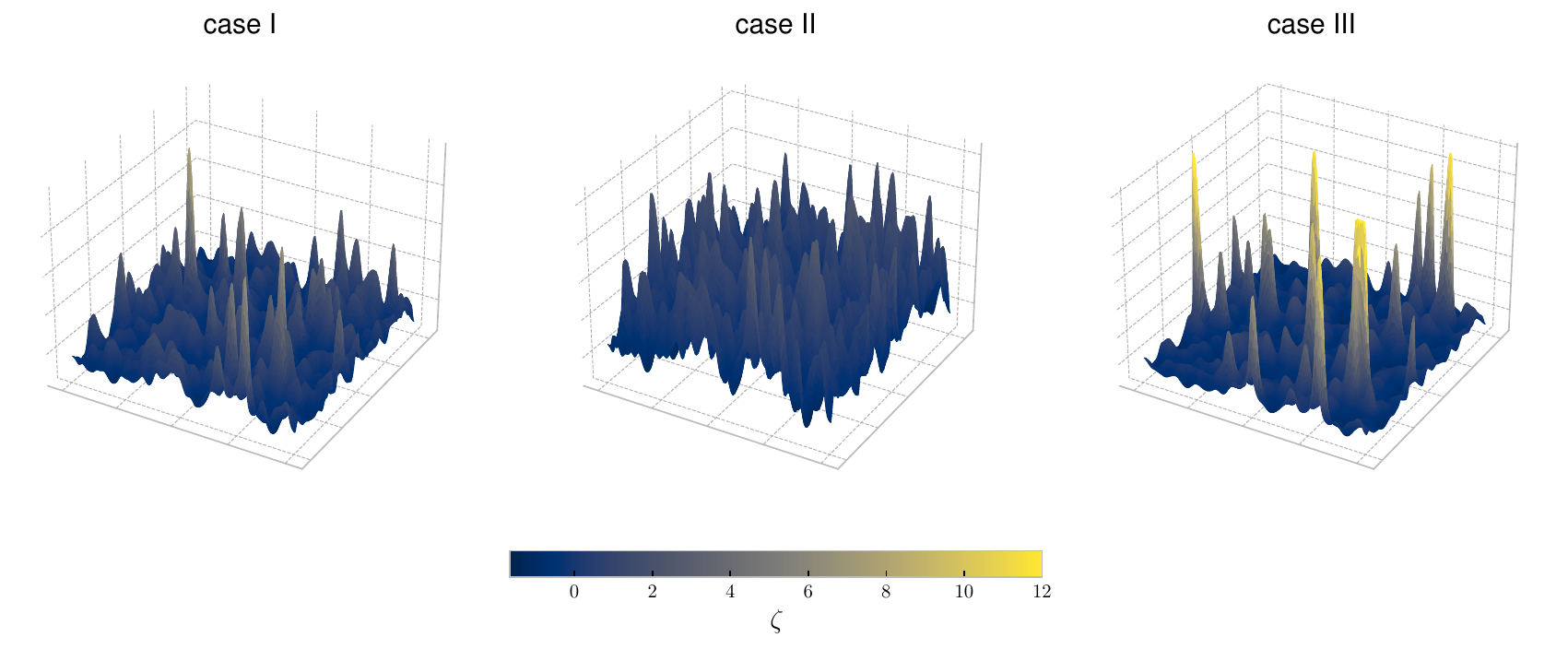}
		\caption{3D surface plots showing the value of $\zeta$ on the same 2D slices as in Figs.~\ref{fig:snapshots} and~\ref{fig:snapshots_negative}, for $\mathcal{P}_{\zeta,\rm tree}^{\rm max} = 1$. The plots display a $100 \times 100$ subset of the $512 \times 512$ simulation snapshots, zoomed in on their center. In the right panel (case III), values of $\zeta>12$ are saturated for visualization purposes. An animated version can be found at the following \href{https://github.com/caravangelo/inflation-easy.git}{link}.}
		\label{fig:3d_snapshots}
	\end{figure*}
	
	In some of the panels in Fig.~\ref{fig:PDF}, we report the value of \(\mathcal{F}\), which represents the fraction of comoving volume contained in the trapped region. This trapping is only relevant in Cases~I and III, which exhibit a local minimum, and is absent in Case~II due to the potential being monotonic. As explained in \cite{Caravano:2024tlp}, \(\mathcal{F}\) provides an estimate of the mass fraction that will collapse into PBHs after inflation ends.\footnote{This population of PBHs is not to be confused with the one potentially produced after reheating from the gravitational collapse of large overdensities when perturbations re-enter the Hubble sphere.}
	
\begin{figure*}
		\centering
		\includegraphics[ width=0.364 \linewidth]{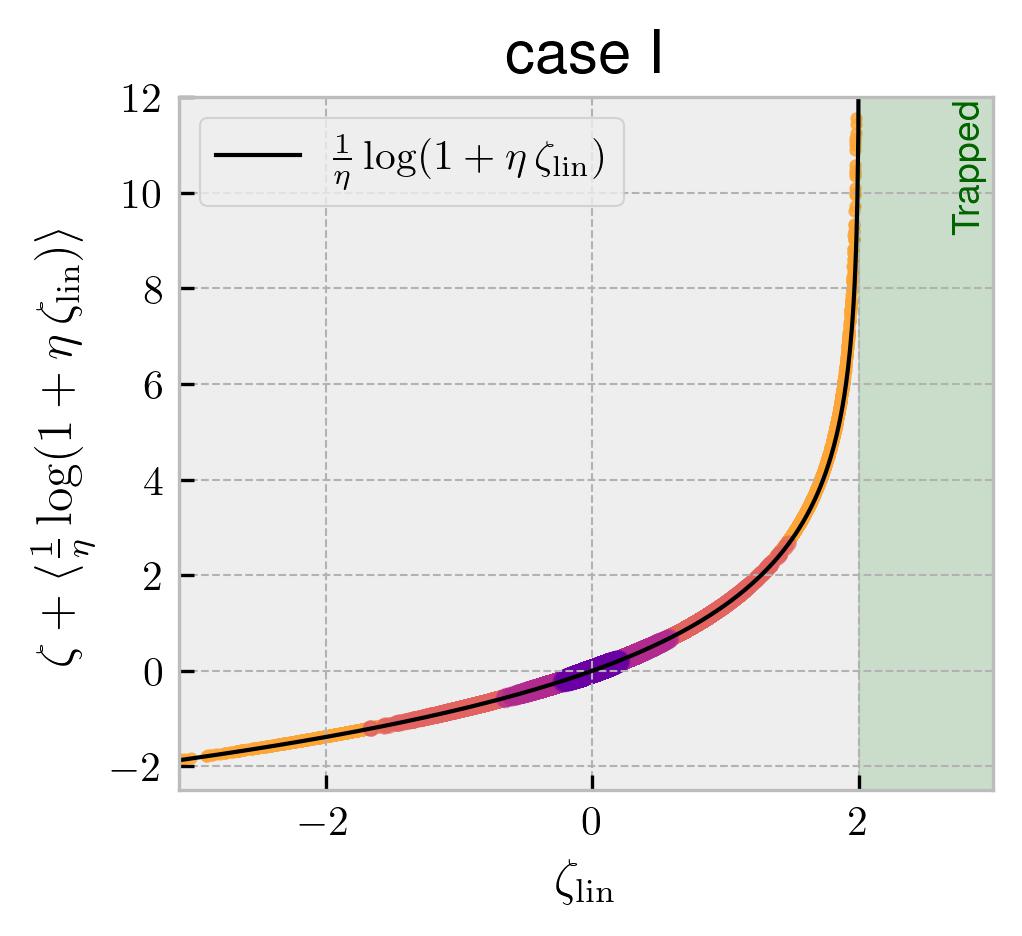}
		\includegraphics[width=0.31 \linewidth]{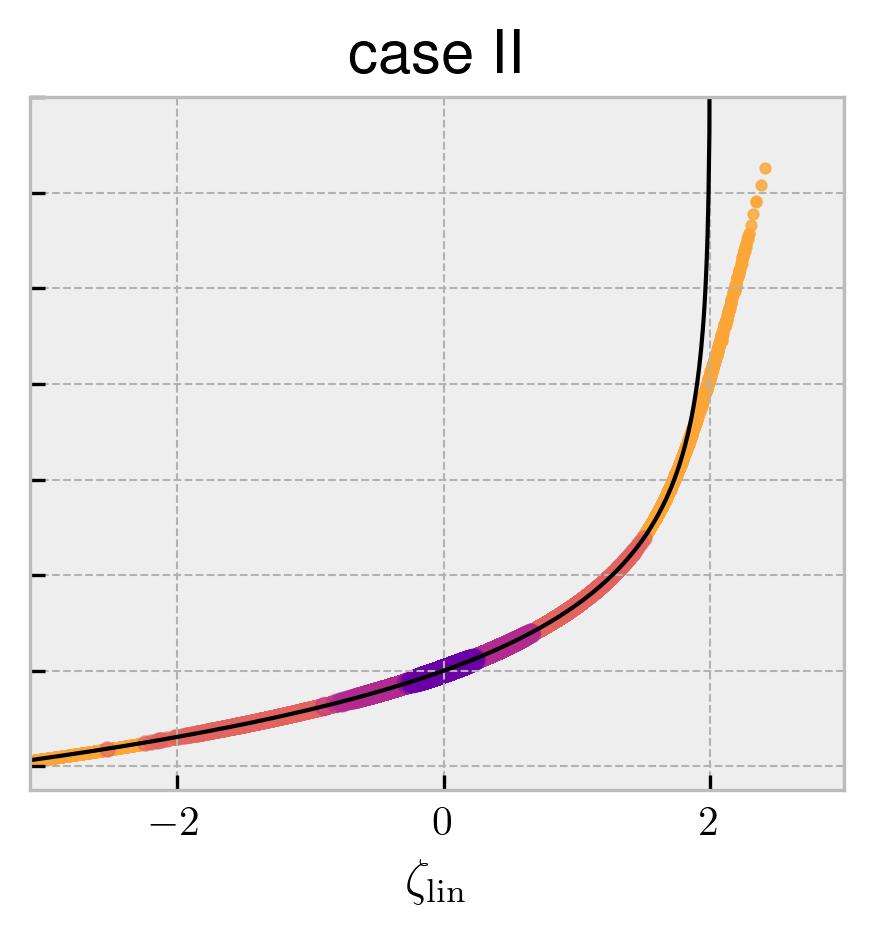}
		\includegraphics[width=0.31 \linewidth]{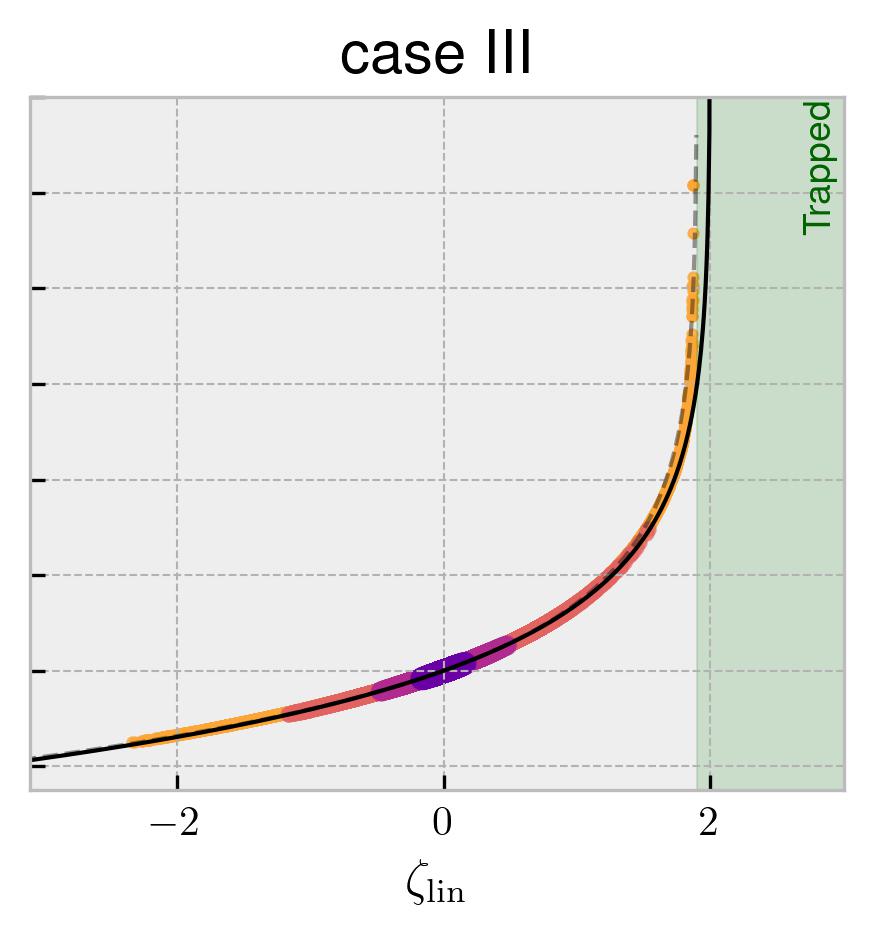}
		\caption{Scatter plot of $\zeta$ vs $\zeta_{\rm lin}=-H\delta\phi/\dot\phi$ from the lattice simulation. The black line shows the log relation \eqref{eq:log} with $\eta=\eta_{\textrm{III}}=-1/2$, while the shaded green indicates regions that are trapped near the local minimum. To the lattice $\zeta$, we have added the mean of the log relation $\langle\zeta_{\rm log}\rangle$, which differs quantitatively from case to case (see the discussion in Sec.~\ref{sec:zero-mode} for details). The dashed black line in the right panel shows the log relation but with $\eta=-1/1.9$ instead of $\eta=\eta_{\textrm{III}}=-1/2$. Different colors correspond to different lattice simulations presented in Paper I, ranging from ${\cal P}_{\zeta,\rm tree}^{\rm  max} =10^{-3}$ to ${\cal P}^{\rm  max}_{\zeta,\rm tree} =1$, where $\cal P_{\zeta,\rm tree}$ is the tree-level power spectrum.}
		\label{fig:scatter}
	\end{figure*}

	Additionally, the vertical gray line in Fig.~\ref{fig:PDF} marks the value \(-H\delta\phi/\dot\phi = -1/\eta_{\textrm{III}} = 2\), where the logarithmic relation in Eq.~\eqref{eq:log} diverges. In case I, characterized by Wands duality, this divergence precisely coincides with the onset of trapping, establishing a one-to-one correspondence between points with divergent $\zeta_{\rm log}$ and the threshold for trapped points. This will be illustrated in Fig.~\ref{fig:scatter}, where we show a plot of \(\zeta\) versus the linear estimate \(\zeta_{\rm lin} = -H\delta\phi/\dot\phi\). 
	This correspondence is broken in the other cases. We will discuss this deviation in more detail in Sec.~\ref{sec:regime-validity-fiducial-background}. 
	
	For additional insight, we examine 2d snapshots of the real-space maps of $\zeta$, in Fig.~\ref{fig:snapshots} for positive \(\zeta\) values and Fig.~\ref{fig:snapshots_negative} for negative values, highlighting the differences between the full nonperturbative \(\zeta\) and $\zeta_{\textrm{lin}}$. Although the nonlinear mapping does not alter the overall structure of the field—determined by the inflationary dynamics around horizon crossing—it significantly amplifies \(\zeta\) in overdense regions, with stronger effects in high-density areas. In these plots, the trapped regions are marked in green. They also show that the trapping at play here is a small-scale phenomenon: by contrast to the global trapping studied in \cite{Caravano:2024tlp} in more inhomogeneous early-universe scenarios, the trapping is only local here, affecting rare regions. 
	
	Before moving to the next section, let us stress again that, when computing both the logarithmic relation and the linear expression \(-H\delta\phi/\dot\phi\), we use the nonperturbative field values obtained from the lattice simulation. These incorporate nonlinear effects in both \(\delta\phi\) and the backreaction on \(\dot\phi\), as demonstrated in Paper I. Since the field morphology is dictated by non-Gaussianities generated at horizon crossing, their role is crucial: if non-Gaussianities were negligible, all three cases would yield the same maps (Figs.~\ref{fig:snapshots}, and~\ref{fig:snapshots_negative}) and distributions (Fig.~\ref{fig:PDF}). 
	These dissimilarities are also particularly evident in Fig.~\ref{fig:3d_snapshots}, where we show a 3D visualization of these snapshots for the case $\mathcal{P}_{\zeta,\rm tree}^{\rm max} = 1$.
	The observed differences across the different cases highlight the importance of field self-interactions captured by our lattice simulations.

	\section{Discussion}
	\label{sec:discussion}

	\begin{figure}
		\centering
		\includegraphics[width=0.45\textwidth]{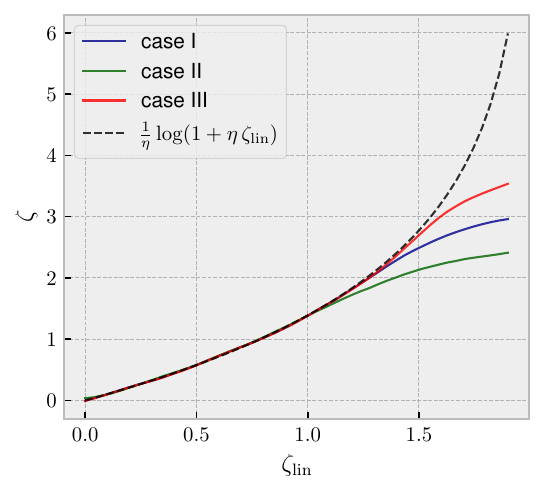}
		\caption{Comparison between $\zeta$ computed from the nonlinear change of gauge \eqref{eq:change-gauge} based on perturbation theory, and the logarithmic relation \eqref{eq:log}, representative of the fully nonperturbative $\delta N$ result in this range, see the main text in Sec.~\ref{sec:quantitative-break-down-PT}. In each case, the scenario with ${\cal P}^{\rm  max}_{\zeta,\rm tree} =1$ is considered.}
		\label{fig:breakdown-PT}
	\end{figure}

	\subsection{Breakdown of the perturbative change of gauge}
	\label{sec:quantitative-break-down-PT}
	
	We explained in Sec.~\ref{sec:change-gauge} that the change of gauge relationship \eqref{eq:change-gauge} is intrinsically perturbative. However, it is hard to know in advance when it provides a good description and when it fails. 
	To gain a deeper understanding of its regime of validity, we compare it to the non-perturbative $\delta N$ approach.
	
	In Fig.~\ref{fig:scatter} we show the relation between  the linear estimate \(\zeta_{\rm lin} = -H\delta\phi/\dot\phi\) and
	\(\zeta\) computed using non-perturbative $\delta N$.
	Colors report the different lattice simulations from Paper~I, spanning ${\cal P}_{\zeta,\rm tree}^{\rm max} = 10^{-3}$ to $1$. 
	The black line shows the log relation~\eqref{eq:log} with $\eta = \eta_{\textrm{III}} = -1/2$, while the green shaded area marks regions trapped near the local minimum. We shift the lattice $\zeta$ by the mean value $\langle\zeta_{\rm log}\rangle$, which varies across cases (see Sec.~\ref{sec:zero-mode}). 
	We see that in case I, the log relation~\eqref{eq:log} agrees with the lattice results in all cases, placing the divergence at $\zeta_{\rm lin} =  - 1/\eta$ due to the Wands duality. 
	The log relation is only a sufficiently good approximation in the other cases when perturbations are not extreme, while it fails for large positive $\zeta_{\rm lin}$ when ${\cal P}^{\rm  max}_{\zeta,\rm tree} =1$ (yellow points). We will elaborate on this in Sec.~\ref{sec:regime-validity-fiducial-background}.

	In Fig.~\ref{fig:breakdown-PT}, we quantify the limitation of the perturbative approach by numerically solving Eq.~\eqref{eq:change-gauge} for the 3 cases under study, in the scenarios with ${\cal P}^{\rm  max}_{\zeta,\rm tree} =1$ that feature the largest fluctuations. This computation takes as an input the background evolution $\phi(N)=\langle \Phi(N,\vec{x}) \rangle$ extracted from the simulations. Hence it can only probe fluctuations that are late in the potential, $\Phi(N,\vec{x}) > \phi(N)$ at the final simulation time, corresponding to $\zeta_\textrm{lin}>0$. This is not a limitation, since the breakdown of the perturbative gauge transformation is most pronounced on this positive branch; in contrast, the negative branch corresponds to trajectories that have reached the constant-$\eta$ attractor.
	
	For simplicity of presentation, we compare the results to the logarithmic relationship \eqref{eq:log}, which, as we have seen in Fig.~\ref{fig:scatter}, is in very good agreement with the nonperturbative $\delta N$ results across all cases for the range of values displayed here. The small offset between $\zeta$ and $\zeta_\textrm{log}$ due to the non-vanishing zero mode $\langle \zeta_\textrm{log} \rangle$, discussed in Sec.~\ref{sec:zero-mode}, does not affect the discussion here. The departure of the perturbative results from the nonperturbative one for $\zeta_\textrm{lin} \gtrsim 1$ is readily apparent. Moreover, the larger values of $\zeta$ compared to $\zetapt$ can be simply understood: such values of $\zeta_\textrm{lin}$ correspond to rare points that are still probing the USR part of the potential, and with much smaller velocities there than the typical behavior described by $\phi(N)$.
	Hence, they take a larger number of $e$-folds to finish inflation compared to what perturbation theory predicts based on the average behavior.
	
	It is worth emphasizing that the departure between $\zeta$ and $\zetapt$ is not uniquely tied to the trapping phenomenon, as it also appears in case II. Rather, it arises from the absence of an attractor during the USR phase, which causes different regions to follow different histories.

	\subsection{Regime of validity of the effective background description}
	\label{sec:regime-validity-fiducial-background}
	
\begin{figure}
		\centering
		\includegraphics[width=0.49\textwidth]{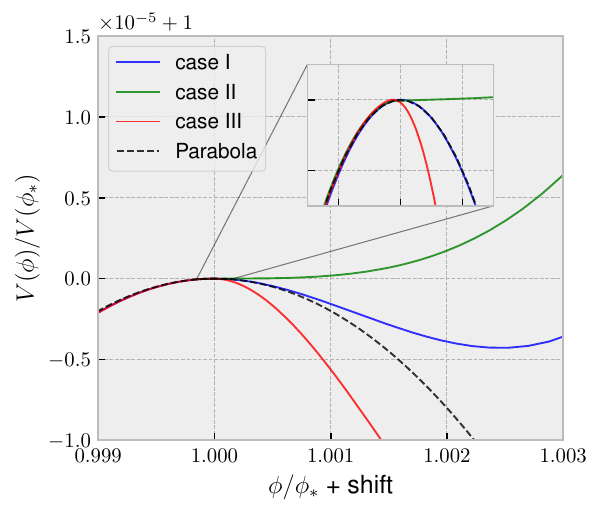}
		\caption{ A zoom-in on the USR potentials, for the three cases for \(\mathcal{P}_{\zeta,\rm tree}^{\rm max} = 1\). In the figure, $\phi_*$ is the field value associated with the local maxima. The parabola, shown as a dashed black line, is defined by the potential in Eq.~\eqref{eq:parabola}.
			Since the absolute value of the field has no physical meaning, we applied an arbitrary shift to the \(x\)-axis to align the shapes of the potentials on the left, corresponding to the second slow-roll attractor.
		}\label{fig:potential_zoom}
	\end{figure}

	The analytical formula \eqref{eq:log} is expected to accurately predict the fully nonperturbative $\zeta$ computed with the $\delta N$ formalism for fluctuations at the end of the simulations that probe the part of the potential corresponding to the dynamical attractor with constant $\eta$. Indeed, for the time range of interest, these points share the same history, following the background $\phi(N)$ that is very well approximated by the effective $\bar{\phi}(N)$, up to a time shift given by $\zeta$. This agreement for the bulk of the distribution is manifest in Figs.~\ref{fig:PDF} and \ref{fig:scatter}.  It is also evident in Fig.~\ref{fig:potential_zoom}, which shows that all three potentials---for the scenarios with \(\mathcal{P}_{\zeta,\rm tree}^{\rm max} = 1\)---closely follow the inverted parabola~\eqref{eq:parabola}, corresponding to constant \(\eta\), in the slow-roll attractor regime to the left of the potential, away from the local maximum. This behavior also explains why the logarithmic formula remains valid even for rare, large fluctuations with \(\zeta_\textrm{lin} < 0\).

	Let us now discuss rare, large fluctuations with $\zeta_\textrm{lin} > 0$. In cases I and III---i.e., for potentials that feature a local minimum and the associated trapping phenomenon---we expect the logarithmic formula \eqref{eq:log} to provide a good estimate of $\zeta$ for large but subcritical fluctuations that do not get trapped. This expectation arises from the construction of the model: such fluctuations reach the local maximum with a very small escape velocity (much smaller than that of typical fluctuations, represented by the background), and their trajectory is therefore well approximated by Eq.~\eqref{eq:fiducial-background}. This is confirmed by the results in Fig.~\ref{fig:scatter}: in contrast to the perturbative change-of-gauge results shown in Fig.~\ref{fig:breakdown-PT}, the logarithmic formula accurately captures the growth of $\zeta$ for large fluctuations in both cases. 
	
	In case~I (WD), this agreement is also strikingly visible in the first panel of Fig.~\ref{fig:PDF}, where the effective background analysis very accurately predicts the entire PDF of \(\zeta\), even in the most inhomogeneous scenario with $\mathcal{P}_{\zeta,\rm tree}^{\rm max} = 1$. This can be understood from the fact that $\nu^2$ remains nearly constant following the onset of USR, or equivalently, that the potential is well approximated by the quadratic form~\eqref{eq:parabola} invariant under Wands duality; see Fig.~\ref{fig:potential_zoom}. With the same potential and the same initial conditions---namely, an asymptotically vanishing velocity near the top of the inverted parabola, as these points lie at the threshold for trapping---the two descriptions yield identical predictions for the duration of inflation in these patches, i.e., for $\zeta$. 
	To summarize the behavior in case~I, although $\bar{\phi}(N)$ differs significantly from the true background $\phi(N)$ away from the constant-\(\eta\) attractor, it interpolates between the two universal behaviors encountered in these models: that of typical fluctuations and that of almost-trapped patches, effectively serving as a common background for both.\footnote{The universality of these behaviors is clearly seen in the scatter plots of Fig.~\ref{fig:scatter}, which in fact show no scatter, indicative of a one-to-one relationship between $\delta\phi$ and $\delta\phi'$, {consistent with the analysis in \cite{Cruces:2025typ} stressing that these are fully correlated on super-Hubble scales.}}
	
	In case~III, by contrast, Figs.~\ref{fig:PDF} and \ref{fig:scatter} show that the effective background description fails to quantitatively predict \(\zeta\) for very large fluctuations close to trapping. This can be understood as follows: although trapped points are released near the top of the potential with almost vanishing velocity, the potential near
	the local maximum differs from the quadratic one \eqref{eq:parabola}, as shown in the zoom-in region of Fig.~\ref{fig:potential_zoom}. Hence, although the duration of inflation also grows without bounds as one considers patches that are closer and closer to being trapped, the trajectory differs from the one computed with the potential \eqref{eq:parabola}. Furthermore, the zoom-in region shows that, while the potentials share the same shape on the second slow-roll attractor at small field values, the local maximum is reached at a smaller field value in case~III than in case~I. This explains why the divergence of $\zeta$ in Fig.~\ref{fig:scatter} occurs at a slightly smaller value of $\zeta_\textrm{lin}$ than $-1/\eta_{\textrm{III}}$. It also turns out that the result is well fitted by the logarithmic relation~\eqref{eq:log} with the value $\eta = -1/1.9$, shown as a dashed black line.
	
	In case~II, the logarithmic formula is not valid for large positive $\zeta_\textrm{lin}$, for obvious reasons: no divergence of $\zeta$ can occur, as there is no trapping mechanism---regardless of the field excursion, the field always reaches the end-of-inflation hypersurface in finite time. In this case, large fluctuations (the points on the right of the scatter plot in Fig.~\ref{fig:scatter}) follow a simple, linear universal behavior. This suggests that $\zeta$ could be derived from an effective background analysis also in this case. The corresponding effective $\bar{\phi}(N)$ would need to reproduce the dynamics of these atypical fluctuations on that part of the potential, and would thus differ from the one introduced in Sec.~\ref{sec:effective-background}.

	Finally, let us remark that the logarithmic formula~\eqref{eq:log} has been frequently used in the literature on USR inflation; see, e.g., \cite{Cai:2018dkf,Biagetti:2018pjj,Atal:2019cdz,Atal:2019erb,Biagetti:2021eep,Pi:2022ysn,Tomberg:2023kli,Ferrante:2022mui,Wang:2024xdl,Ballesteros:2024pwn,Inui:2024sce,Inui:2024fgk,Ballesteros:2024pbe}. Our use of it differs from most studies in three ways. First, it provides an interesting analytical recipe but primarily serves as a comparison to the exact \(\delta N\) computation performed using the lattice simulation. Second, this relation has been predominantly applied assuming Gaussian statistics for $\zeta_{\textrm{lin}}$, which does not hold for $\zeta_{\textrm{lin}}$ obtained from the lattice simulations, whose statistics can be strongly non-Gaussian in general.\footnote{See \cite{Ballesteros:2024pbe} for a notable study applying \eqref{eq:log} to realizations encoding the perturbative computation of the bispectrum of \(\zeta_{\textrm{lin}}\).} Third, to the best of our knowledge, the extension of perturbative change-of-gauge methods to handle almost trapped patches via an effective background description, as presented here, is novel.

	\subsection{Nonperturbative $\zeta$, perturbative $\zeta$, and its zero mode}
	\label{sec:zero-mode}

	We have seen that the curvature perturbation deduced from the change of gauge reasoning (see Eq.~\eqref{eq:change-gauge}) is intrinsically perturbative. Hence, it is such that $\delta \phi=0$ corresponds to $\zetapt=0$, i.e.~to an expansion $\zetapt=-\delta \phi/\phi'+ \ldots$. By contrast, the nonperturbative $\zeta$ in Eq.~\eqref{eq:zeta-delta-N} is defined such that its spatial average vanishes. The two are thus distinct a priori: there is no reason that points with $\delta \phi=0$ generate the same number of $e$-folds as the average $\langle N(\vec{x}) \rangle$ in Eq.~\eqref{eq:zeta-delta-N}. Or, equivalently, $\langle \zetapt \rangle$ does not vanish in general, i.e.~it has a zero mode that generates an offset between $\zeta$ and $\zetapt$, even for small fluctuations for which the perturbative change of gauge correctly captures the physics.
	
	Naturally, this discussion also applies to the curvature perturbation \eqref{eq:log} deduced from the effective background description, which results from a change of gauge applied to a specific background. This is why, in Fig.~\ref{fig:scatter}, we compared $\zeta_\textrm{log}$ and $\zeta+\langle \zeta_\textrm{log} \rangle$, to allow a meaningful comparison. 
	
\begin{figure}
		\centering
		\includegraphics[width=0.49\textwidth]{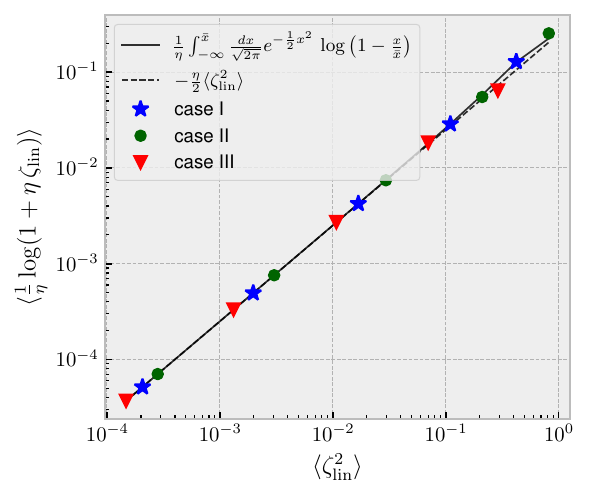}
		\caption{Plot of the zero mode $\langle\zeta_{\rm log}\rangle$ across all cases and scenarios. The horizontal axis represents the variance $\langle\zeta_{\rm lin}^2\rangle$ obtained from the lattice simulations. The black line shows the analytical result \eqref{eq:average-zeta-log} corresponding to a Gaussian statistics for $\zeta_{\rm lin}$, which is also compared to the approximated relation \eqref{eq:average-zeta-log-approximate}.}\label{fig:zero_mode}
	\end{figure}

	This zero mode does not impact observables for our purposes, but it is interesting to quantify, which we do in Fig.~\ref{fig:zero_mode}, where we show results for $\langle \zeta_\textrm{log} \rangle$ computed from the simulations across all cases and scenarios. The black curve corresponds to the result derived by assuming that $\zeta_{\rm lin}$ obeys Gaussian statistics, in which case 
	\begin{equation}
		\langle \zeta_{\rm log}\rangle = \frac{1}{\eta} \int_{-\infty}^{\bar{x}} \frac{\mathrm{d}x}{\sqrt{2\pi}}\, e^{-\frac{1}{2}x^2} \log\left(1 - \frac{x}{\bar{x}}\right),
		\label{eq:average-zeta-log}
	\end{equation}
	where $\bar{x} = -(\eta\, \langle \zeta_{\rm lin}^2 \rangle^{1/2})^{-1}$.

	This integral can be performed analytically, but its explicit expression in terms of hypergeometric functions is not illuminating. However, the result can be easily understood: $\bar{x}$ measures the scale $-1/\eta$ entering into the logarithmic relation in units of the typical range of values of $\zeta_\textrm{lin}$. Hence, for large $\bar{x}$, the bulk of the integral comes from the region around $x=0$, and one can effectively expand the logarithmic term in powers of $x/\bar{x}$ and extend the integral up to infinity, in which case one finds
	\begin{equation}
		\langle \zeta_{\rm log}\rangle \simeq -\frac{\eta}{2} \langle \zeta_{\rm lin}^2 \rangle\,,
		\label{eq:average-zeta-log-approximate}
	\end{equation}
	recovering the expectation of perturbation theory that the offset goes to zero as the variance decreases. In Fig.~\ref{fig:zero_mode}, one sees that this simple expression is actually a very good approximation to the analytical result up to values $\langle \zeta_{\rm lin}^2 \rangle \approx 1$.

	Not surprisingly, the analytical Gaussian result agrees perfectly with the lattice results in case~I, characterized by Wands duality and a Gaussian distribution of $\zeta_{\rm lin}$. Remarkably, this simple result also reproduces the lattice results in cases~II and~III for which $\zeta_{\rm lin}$ has strong non-Gaussian tails. This has a simple explanation: most of the contribution to $\langle \zeta_{\rm log} \rangle$ comes from typical fluctuations, where the probability density function of $\zeta_{\rm lin}$ can be well approximated by a Gaussian, see Fig.~\ref{fig:PDF}.
	
	We further observe in Fig.~\ref{fig:zero_mode} that, for fixed values of the tree-level power spectrum, the value of $\langle \zeta_{\rm lin}^2 \rangle$ in case~III is always lower than in case~I, which in turn is lower than in case~II. This pattern causes the three colors---red, green, and blue---to alternate accordingly in the figure. This behavior reflects the broadness of the peak in the power spectrum of $\zeta_{\rm lin}$, as shown in Fig.~\ref{fig:PS}. Specifically, case~II features a plateau-like structure, whereas cases~I and~III exhibit sharply peaked power spectra, with case~III being the most sharply peaked. As a result, the overall integrated power is smaller in case~III, and higher in case~II, compared to case~I.
	
	We conclude this discussion with a final remark. The ``zero mode'' referred to in this section denotes a constant average quantity within the lattice box and is therefore only meaningful for scales contained in the box and thus close to the peak of the power spectrum. Importantly, it cannot be directly related to the ongoing debate concerning the connection between the USR phase and loop corrections to large-scale (i.e., CMB) modes (e.g.  \cite{Kristiano:2022maq,Cheng:2021lif,Riotto:2023hoz,Kristiano:2023scm,Riotto:2023gpm,Kristiano:2024vst,Franciolini:2023lgy, Davies:2023hhn,Firouzjahi:2023ahg, Iacconi:2023ggt,Motohashi:2023syh,Tasinato:2023ioq, Tasinato:2023ukp,Firouzjahi:2023aum,Fumagalli:2023hpa,Cheng:2023ikq,Fumagalli:2023loc,Tada:2023rgp,Firouzjahi:2023bkt,Braglia:2024zsl, Kawaguchi:2024lsw,Ballesteros:2024zdp,Inomata:2024lud,Fumagalli:2024jzz,Kawaguchi:2024rsv,Green:2024fsz,Inomata:2025bqw,Fang:2025vhi,Inomata:2025pqa,Braglia:2025cee,Kristiano:2025ajj,Braglia:2025qrb}), for several reasons. First, our setup does not access such large scales, and we cannot determine whether this ``zero'' mode would manifest coherently on super-lattice scales—as required for it to be interpreted as a perturbation—or whether it would be volume suppressed when averaging over multiple patches of lattice size. Second, if this mode persists as a deterministic constant shift across all scales, it would correspond to a renormalization of the scale factor, effectively acting as a backreaction on the background evolution, and would thus have no observable impact on late-time cosmological measurements.
	
	\section{Conclusions}\label{sec:conclusions}

	In this work, we extended the analysis of Paper I~\cite{Caravano:2024moy} by computing the curvature perturbation $\zeta$ directly from lattice simulations of USR dynamics. To extract $\zeta$ in a fully nonperturbative manner, we combined the lattice output with the $\delta N$ formalism, which identifies $\zeta$ as the local number of $e$-folds required to reach a uniform-density hypersurface from a spatially flat one. Applied to the final field configuration from the lattice, this method captures all sources of nonlinearity: the backreaction of fluctuations on the background dynamics, the intrinsic non-Gaussianity generated around horizon crossing due to nonlinear field evolution, and the gravitational nonlinearity in the mapping between inflaton fluctuations and $\zeta$. The intrinsic non-Gaussianity includes effects from attractive or repulsive self-interactions of the field—discussed in detail in Paper I—that are neglected in stochastic approaches.
	
	A central goal of this paper was to isolate the impact of the nonlinear mapping between $\phi$ and $\zeta$. To this end, we compared the fully nonperturbative $\zeta$ field obtained via the $\delta N$ method with a “linear” curvature perturbation, $\zeta_{\rm lin} = -H\delta\phi / \dot\phi$, constructed by applying the linear gauge relation to the $\delta\phi$ and $\dot\phi$ extracted from the lattice simulations. Since both $\zeta$ and $\zeta_{\textrm{lin}}$ are derived from the same lattice field configuration, any difference between them isolates the effect of the nonlinear mapping itself.
	
	We found that this nonlinear relation introduces significant modifications to the curvature perturbation. While its effect on the power spectrum is relatively modest—particularly compared to the larger corrections due to backreaction during the USR dynamics, as discussed in Paper I—it has a pronounced impact on higher-order statistics. Specifically, it substantially enhances the positive tail of the probability distribution of $\zeta$. This effect is clearly visible in real-space snapshots as a marked amplification of overdense regions.
	
	Another central focus of this work was comparing our results with the standard perturbative approach, in which the relation between $\delta\phi$ and $\zeta$ is treated as a gauge transformation. To rigorously assess the breakdown of this perturbative picture, we revisited the change-of-gauge argument and highlighted its underlying assumptions. We applied this reasoning to an effective background with constant $\eta$, which leads to an analytic nonlinear relation between $\delta\phi$ and $\zeta$: the logarithmic form in Eq.~\eqref{eq:log}. It is important to stress that despite being nonlinear, this relation remains perturbative, as it relies on the assumption that all space-time points follow a single, well-defined background trajectory. Because our model enters a constant-$\eta$ phase after the USR period, this effective background approximation reproduces the nonperturbative results with excellent accuracy as long as fluctuations remain small.
	
	Comparing our results with a perturbative framework---both for the actual background and for the effective background description giving rise to the logarithmic relation---
	proved extremely valuable, as it allowed us to provide a clear picture of the breakdown of perturbative theory for large fluctuations. This breakdown is due to regions of space following a history distinct from the background. In cases I and III, this can lead to a trapping phenomenon, in which certain patches never end inflation.
	In Case II, which lacks a local minimum in the potential and therefore exhibits no trapping, the breakdown arises because large fluctuations can remain in the USR phase with a velocity much smaller than the background, while most of the universe has already transitioned to the slow-roll attractor. 
	
	Interestingly, in Case I---where the model is approximately free due to Wands duality---the fiducial background trajectory underlying the logarithmic relation still provides a good description even for regions near the trapping threshold, which can exhibit arbitrarily large values of $\zeta$. This is a special feature of the model, which was engineered to mimic an inverted parabola all the way up to the local maximum, as discussed in Sec.~\ref{sec:discussion}. In Cases II and III, this is not the case: for large fluctuations, the logarithmic relation fails to provide a correct value of $\zeta$, significantly overestimating the tail of the probability distribution, and thereby the predicted abundance of PBHs.

	We have also demonstrated that the non-linear mapping between $\delta\phi$ fluctuations and the curvature perturbation $\zeta$ generically induces a positive skewness in the PDF of $\zeta$, even in Case II, which is characterized by ``negative'' NGs (see Paper I). This finding supports and extends the argument put forward in Ref.~\cite{Firouzjahi:2023xke}, which suggested that USR scenarios yield a positive $f_{\rm NL}$ near the peak scale. As a result, PBH formation in USR models is expected to be enhanced relative to naive Gaussian estimates. These results have important implications for interpreting GW observations \cite{NANOGrav:2023gor,EPTA:2023fyk,Reardon:2023gzh,Xu:2023wog,Romero-Rodriguez:2021aws,LISACosmologyWorkingGroup:2024hsc,LISACosmologyWorkingGroup:2025vdz}, particularly in scenarios requiring large amplitudes near the PBH overproduction threshold, as may be the case for recent pulsar timing array signals \cite{EPTA:2023xxk,NANOGrav:2023hvm,Franciolini:2023pbf,Figueroa:2023zhu,Ellis:2023oxs,Balaji:2023ehk,Domenech:2024rks,Iovino:2024tyg,Cecchini:2025oks}.

	Finally, we emphasize that an accurate prediction of PBH abundance relies on determining the spatial profile of the collapsing overdensities, which is necessary to identify the threshold for collapse from relativistic numerical simulations \cite{Musco:2018rwt,Escriva:2019phb,Musco:2020jjb}. 
	Although previous studies using the analytical log mapping estimated the profile by applying the transformation to the peak profile of a Gaussian field (see e.g.~\cite{Yoo:2019pma,Kitajima:2021fpq,Inui:2024fgk}), we expect the final profile to be modified by the intrinsic nonlinearity captured by the lattice simulation.
	This has significant consequences for PBH production, which is exponentially sensitive to the threshold \cite{Young:2024jsu}. This information can, in principle, be extracted directly from the simulations, underscoring the importance of further developing lattice-based methods within this context.
	
	Together with Paper I, this work provides a complete nonlinear and nonperturbative characterization of USR inflationary dynamics. While our analysis focused on a single-field USR scenario, the methodology developed here is broadly applicable and can be extended to more general inflationary models, including multi-field setups. Although we have concentrated on computing the curvature perturbation, the same framework could be extended to directly extract measurable quantities, such as the abundance of PBHs and the spectrum of SIGWs. This will require extending the simulation infrastructure to include the tensor sector and modeling the post-inflationary evolution of perturbations. We leave these exciting developments to future work.
	
	More broadly, this study highlights the power of lattice simulations as a unique tool for probing the nonlinear dynamics of inflation. As this work demonstrates, they offer a self-consistent and fully nonperturbative framework that can capture phenomena far beyond the reach of conventional perturbative techniques. With the next generation of cosmological experiments soon able to test inflationary predictions with unprecedented precision, the need for robust, first-principles predictions---especially for non-Gaussianity, gravitational waves, and rare event statistics---has become increasingly urgent. Lattice simulations may play a central role in bridging the gap between inflationary theory and observations, improving the testability of the physics at play in the very early universe physics. The code used in this work is publicly available at the following \href{https://github.com/caravangelo/inflation-easy.git}{link}.

	\let\oldaddcontentsline\addcontentsline% Store \addcontentsline
	\renewcommand{\addcontentsline}[3]{}% Make \addcontentsline a no-op
	\begin{acknowledgments}
		We thank Laura Iacconi, Antonio Riotto, and David Wands for useful discussions. A.C. acknowledges funding support from the Initiative Physique des Infinis (IPI), a research training program of the Idex SUPER at Sorbonne Universit\'e. This article is distributed under the Creative Commons Attribution International Licence
		(\href{https://creativecommons.org/licenses/by/4.0/}{CC-BY 4.0}).
	\end{acknowledgments}
	
	\twocolumngrid
	\bibliography{main}
	\let\addcontentsline\oldaddcontentsline% Restore \addcontentsline
	
\end{document}